\documentclass{aa}

\usepackage{graphicx}
\usepackage[varg]{txfonts}
\usepackage{bm}

\newcommand{\beq}{\begin{equation}}
\newcommand{\eeq}{\end{equation}}
\newcommand{\bea}{\begin{eqnarray}}
\newcommand{\eea}{\end{eqnarray}}
\newcommand{\req}[1]{Eq.~(\ref{#1})}

\newcommand{\dd}{\mathrm{d}}
\newcommand{\erm}[1]{\mathrm{e}^{#1}}
\newcommand{\gcc}{\mbox{g cm$^{-3}$}}
\newcommand{\kB}{k_\mathrm{B}}
\newcommand{\mel}{m_\mathrm{e}}
\newcommand{\nel}{n_\mathrm{e}}
\newcommand{\nbar}{\bar{n}}
\newcommand{\nion}{n_\mathrm{i}}

\newcommand{\npd}{n_\mathrm{pd}}

\newcommand{\rg}{r_\mathrm{g}}
\newcommand{\rhob}{\rho_\mathrm{b}}
\newcommand{\Ta}{T_*}

\newcommand{\Teff}{T_\mathrm{eff}}
\newcommand{\tTeff}{\tilde{T}_\mathrm{eff}}
\newcommand{\zg}{z_\mathrm{g}}

\begin{document}

\title{Crust structure and thermal evolution of neutron stars
\\
 in soft X-ray transients}
                                                         
\author{
A. Y. Potekhin\inst{1,2}\thanks{\email{palex@astro.ioffe.ru}}
\and
G. Chabrier\inst{1,3}
}
\institute{
Ecole Normale
Sup\'erieure de Lyon, CRAL (CNRS, UMR 5574), 46 all\'ee d'Italie,
69364, Lyon Cedex 07, France
\and
Ioffe Institute,
Politekhnicheskaya 26, 194021, Saint Petersburg, Russia
\and
School of Physics, University of Exeter, Exeter, UK EX4 4QL
}

\date{Received 23 July 2020 / Accepted 16 November 2020}

%%%%%%%%%%%%%%%%%%%%%%%%%%%%%%%%%%%%%%%%%%%%%%%%%%%%%%%%%
\abstract
{Thermal evolution of neutron stars in soft X-ray transients (SXTs) is
sensitive to the equation of state, nucleon superfluidity,
composition and structure of the crust.  Comparison of observations of
their crust cooling with simulations is a powerful tool of verification
of theoretical models of the dense matter.
}{We study the effect of physics input on thermal evolution of neutron
stars in SXTs. In particular, we consider different modern models of the
sources of deep crustal heating during accretion episodes and the
effects brought about by impurities embedded in the crust during its
formation.
}{We simulate thermal structure and evolution of episodically accreting
neutron stars under different assumptions on the crust composition and
on the distribution of heat sources and impurities. For the
nonaccreted crust, we consider  the nuclear charge fluctuations that
arise at crust formation. For the accreted crust, we compare different
theoretical models of composition and internal heating. We 
also compare
results of numerical simulations with observations of the
crust cooling in SXT MXB 1659$-$29.
}{The nonaccreted part of the inner crust of a neutron star can have a
layered structure, with almost pure crystalline layers interchanging
with layers composed of mixtures of different nuclei. The latter layers
have relatively low thermal conductivities, which affects thermal
evolution of the transients. The impurity distribution in the crust
strongly depends on the models of the dense matter and the crust
formation scenario. The shallow heating that is needed  to reach
agreement between the theory and observations depends on
characteristics of the crust and envelope.}{}


\maketitle

%%%%%%%%%%%%%%%%%%%%%%%%%%%%%%%%%%
\section{Introduction}
\label{sect:intro}

Neutron stars are the most compact stars ever observed: with typical
masses $M\sim 1$\,--\,$2\, M_\odot$, where $M_\odot\approx
2\times10^{33}$~g is the solar mass, they have radii $R\approx10-14$ km.
Comparison of observed properties of these stars with theoretical
predictions can provide information on the poorly known properties of
ultra-dense matter in their interiors.

Many neutron stars reside in binary systems with a lower-mass companion
star (low-mass X-ray binaries, LMXBs) and accrete matter
from the companion. Some of the LMXBs, called soft X-ray
transients (SXTs), alternate between phases of accretion (outbursts) and
periods of quiescence. During an outburst, the LMXB emission is
dominated by the accretion disk or a boundary layer (e.g.,
\citealt{InogamovSunyaev10}, and references therein). The accreted
material is fused into heavier elements in the ocean below the boundary
layer, producing heat that can raise the ocean temperature well above
the equilibrium \citep{Fujimoto_84,Fujimoto_87}.  The released 
gravitational energy is so high that X-ray  luminosity reaches $\sim
(10^{36}-10^{38})$ erg~s$^{-1}$. In quiescence, the accretion is
switched off or strongly suppressed, and the luminosity decreases by
several orders of magnitude (see, e.g.,
\citealt{WijnandsDP17} for a review).

When the accreted matter falls onto the neutron star, it pushes the
underlying matter down to denser layers, where electron captures,
neutron emission, and pycnonuclear reactions result in the \emph{deep
crustal heating}. The original ``catalyzed'' crust is gradually replaced
by a crust composed of accreted matter. Once an SXT turns to quiescence,
thermal X-ray emission originates from the surface of the neutron star.
Some of such systems, so called quasi-persistent SXTs, have long
outbursts (lasting months or years), sufficient to appreciably warm up
their crust, and still longer periods of quiescence, when the thermal
relaxation of the overheated crust can be observed directly (e.g.,
\citealt{WijnandsDP17} and references therein). 

{By an analysis of observations of the post-outburst cooling, one
can constrain the thermal conductivity and heat capacity of the crust
(e.g., \citealt{Rutledge_02,Shternin_07,PageReddy13}). In particular,
\citet{PageReddy13} have shown that such constraints, based on
observations of thermal relaxation during a few years, can help to
eliminate much of the uncertainty for an analysis of longer term
variability of neutron star thermal luminosity controlled by the neutron
star core temperature. However, such an analysis can be complicated.
Lightcurves of some SXTs in quiescence can be reproduced within the deep
crustal heating scenario, but require so called \emph{shallow heating},
that is some additional energy sources at relatively low densities
\citep{BrownCumming09}. Some other SXTs can only be explained with
models beyond crustal cooling, involving such processes as residual
accretion during quiescence \citep{TurlioneAP15}.
}

In the previous paper (\citealt{PCC19};
hereafter, Paper~I), we studied the
long-term evolution of the neutron stars in the SXTs, which determines
the equilibrium level of the quiescent emission. Here we study the
relatively short-term thermal evolution of the neutron stars in the
quasi-persistent SXTs during and between accretion episodes.

In Sect.~\ref{sect:CKY} we consider the conventional model, which
assumes that the crust consists of a
sequence of layers, each containing only a single species of nuclei. We
study the influence of several model assumptions, employed in recent
literature, on thermal structure and post-outburst relaxation of the
SXTs. In particular, we consider the effects of simplifying assumptions
about the microphysics of the crust and the distribution of the heat
sources inside the crust on the thermal structure and evolution of the
quasi-persistent SXTs.

{Based on numerical microscopic simulations, \citet{Horowitz_15}
concluded that the neutron star mantle, that is a layer of
nonspherical nuclei (``nuclear pasta''), which is predicted by some
theoretical models at the densities near the
crust-core transition, can have low electrical and thermal
conductivities. However, \citet{NandiSchramm18} found that the structure
factors in the pasta phase are similar to those
in the ordinary phase of quasi-spherical nuclei, and argued that 
the conductivities should be consequently also similar.}
In Sect.~\ref{sect:imp} we 
 note that 
 {the ordinary phase can also be highly resistive} 
 at high densities deep in the
crust. 
{At such densities,}
 the energy differences between different nuclei may become
comparable with the thermal energy of a few hundred keV, at which the
crust is forged. Then the statistical equilibrium allows a mixture of
different nuclei, which is preserved as the subsequent cooling quenches
rapid thermonuclear reactions. We call it \emph{frozen equilibrium
composition} for short. It results in lowering electrical and thermal
conductivities inside the inner crust of the neutron star. 

We adopt the nuclear energy dependences on the charge number $Z$ and on
the mean baryon number density $\nbar$ that were published by
\citet{Pearson_18}, who considered four theoretical models of nuclear
matter from the Brussels-Skyrme family of energy-density functionals.
The first three of them, BSk22, BSk24, and BSk25, behave similarly at
supranuclear densities, being all adjusted to the relatively stiff
microscopic EoS V18 of \citet{LiSchulze08}, although they differ by the
value of the nuclear symmetry energy $J$. The fourth model, BSk26, is
fitted to a softer microscopic EoS of \citet{APR98}. We evaluate the
frozen-equilibrium composition of the crust, calculate the electrical and
thermal conductivities, and use them in the simulations of the
episodically accreting neutron stars.

In Sect.~\ref{sect:MXB} we apply different crust models to simulations
of the thermal history of the neutron star in quasi-persistent SXT MXB
1659$-$29. We perform self-consistent numerical simulations of the
long-term and short-term thermal evolution of this neutron star using
two of the above-mentioned  EoS models, BSk24 and BSk25, which give
similar core compositions and mass-radius relations, but sharply
different frozen-equilibrium mixtures in the deep layers of the inner
crust. Following previous studies of this SXT
\citep{BrownCumming09,Cackett_13,Deibel_17,Parikh_19}, we include 
shallow heating during accretion and a charge impurity parameter
of the accreted crust as additional adjustable model parameters. We 
examine and discuss the influence of the model parameters on the
comparison of the theory with observations.

Summary and outlook are given in Sect.~\ref{sect:concl}.

%%%%%%%%%%%%%%%%%%%%%%%%%%%%%%%%%%
\section{The effects of deep crustal heating models}
\label{sect:CKY}

During accretion, the envelopes, ocean, and crust are gradually
replaced by fresh material. In the outer envelopes, up to the density
$\rho\sim10^8-10^9$ \gcc, the initial iron-group element composition is
replaced by the material of the outer layers of the companion star or by
the products of its thermonuclear burning (see \citealt{Meisel_18} for
review). Deeper in the crust, accreted matter is reprocessed by electron
captures, neutron emissions, and pycnonuclear reactions. The primary
goal of this section is to check the effects of different approximations
to the deep crustal heating on the thermal evolution of a neutron star
during and after an outburst.

Practical models of an accreted crust have been developed by
\citet{HZ90,HZ03,HZ08}, based on the compressible liquid drop model by
\citet{MackieBaym77}. For numerical simulations in this work, we select
the version of the accreted-crust composition and respective energy
releases at the boundaries of different layers that is given in
Table~A.3 of \citet{HZ08} (hereafter HZ'08).

For $T\lesssim3\times10^9$~K, nuclear shell and pairing effects
appreciably affect the nuclear composition of the crust. The role of
these effects in the formation of the accreted crust was studied by
\citet{Fantina_18}, who also presented several practical models for the
crust composition and deep heating. For numerical simulations, we choose
the version given by Table~A.1 of \citet{Fantina_18} (hereafter F+18).

%%%%%%%%%%%%%%%%%%%%%%%%%%%%%%%%%%%%%%%%%%%%%%%%%%%%%%%%%
\begin{figure}
\centering
\includegraphics[width=\columnwidth]{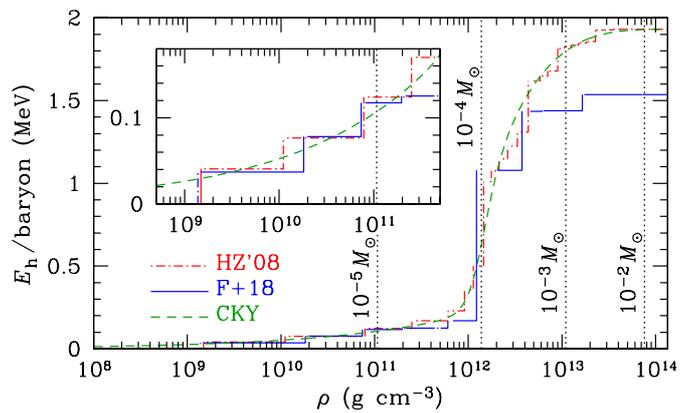}
\caption{Total heat $E_\mathrm{h}$, generated per an accreted baryon  in
a layer from the star surface to a given density, as function of mass
density $\rho$, according to the models of \citet{HZ08} (HZ'08,
dot-dashed red line) and \citet{Fantina_18} (F+18, solid blue line). The
gaps in the lines correspond to the density discontinuities at the phase
boundaries. The dashed green line without gaps shows a smooth analytical
version of the model HZ'08, suggested by \citet{CKY} (CKY). The vertical
dotted lines mark the $\rho$ values corresponding to four masses of
accreted material, from $10^{-5}\,M_\odot$ to $10^{-2}\,M_\odot$,
labeled near these lines, for a neutron star with gravitational mass
$M=1.4\,M_\odot$ and radius $R=12.6$~km.  The inset shows a zoom to the
low-density region.
}
\label{fig:heatsrc}
\end{figure}
%%%%%%%%%%%%%%%%%%%%%%%%%%%%%%%%%%%%%%%%%%%%%%%%%%%%%%%%%

In all above-mentioned models, the nuclear transformations in the course
of the accretion occur at fixed pressures. The  corresponding heat
sources are concentrated at spherical shells.  \citet{CKY} (hereafter
CKY) studied neutron star crust cooling using a smooth analytical
approximation to the HZ'08 model.  Thermal evolution was modeled for
$\rho>\rhob=10^9$ \gcc. The heat flux at the outer boundary $\rho=\rhob$
was linked to the effective surface temperature assuming a
quasi-stationary blanketing envelope (i.e., instantaneous heat
transfer), composed of  light elements.

The model HZ'08 and its smooth approximation CKY predict a total release
of $E_\mathrm{h}=1.93$ MeV of heat per accreted baryon, and the F+18
model predicts $E_\mathrm{h}=1.54$ MeV per baryon.
Figure~\ref{fig:heatsrc} displays the total heat generated per accreted
baryon, from the surface to a given density in the crust, as function of
mass density, for the models HZ'08, CKY, and F+18.

%%%%%%%%%%%%%%%%%%%%%%%%%%%%%%%%%%%%%%%%%%%%%%%%%%%%%%%%%
\begin{figure}
\centering
\includegraphics[width=\columnwidth]{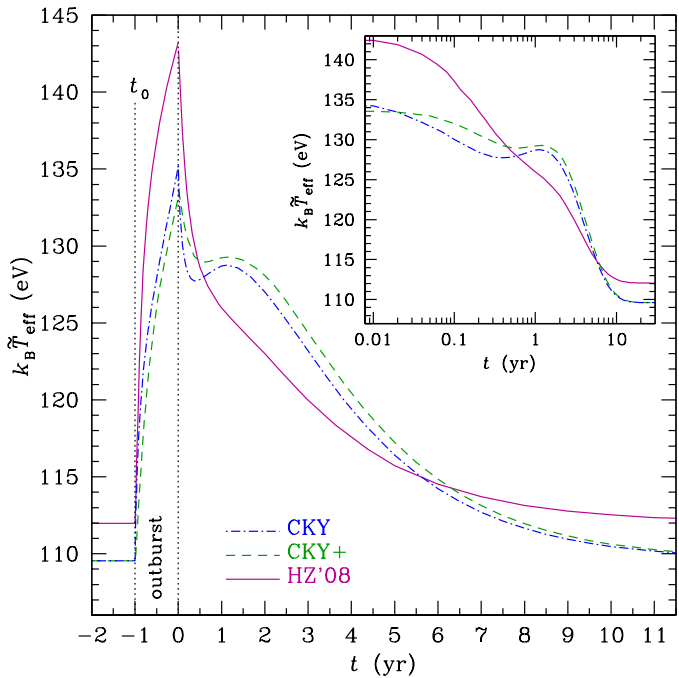}
\caption{Lightcurves for a ``strong outburst'' model of CKY (see text),
computed for different crust models. The heat flux toward the surface of
a neutron star is expressed in terms of redshifted effective temperature
$\tTeff$ and plotted as function of time measured from the end of the
outburst. The dot-dashed line corresponds to the original CKY model,
viz.{} the helium quasi-stationary envelope at $\rho<10^9$ \gcc{} and
the ground-state composition at higher densities with  the heat sources
distributed according to the CKY approximation. The dashed line (CKY+)
shows the same model, but with an accurate treatment of thermal
evolution of the envelope (beyond the approximation of a
quasi-stationary envelope). The solid line represents a more accurate
treatment of the crust with composition and heat source distribution
according to the HZ'08 model. The inset shows the same three curves in
logarithmic scale for time after the end of the outburst, $t>0$.
}
\label{fig:tcky}
\end{figure}
%%%%%%%%%%%%%%%%%%%%%%%%%%%%%%%%%%%%%%%%%%%%%%%%%%%%%%%%%

%%%%%%%%%%%%%%%%%%%%%%%%%%%%%%%%%%%%%%%%%%%%%%%%%%%%%%%%%
\begin{figure}
\centering
\includegraphics[width=\columnwidth]{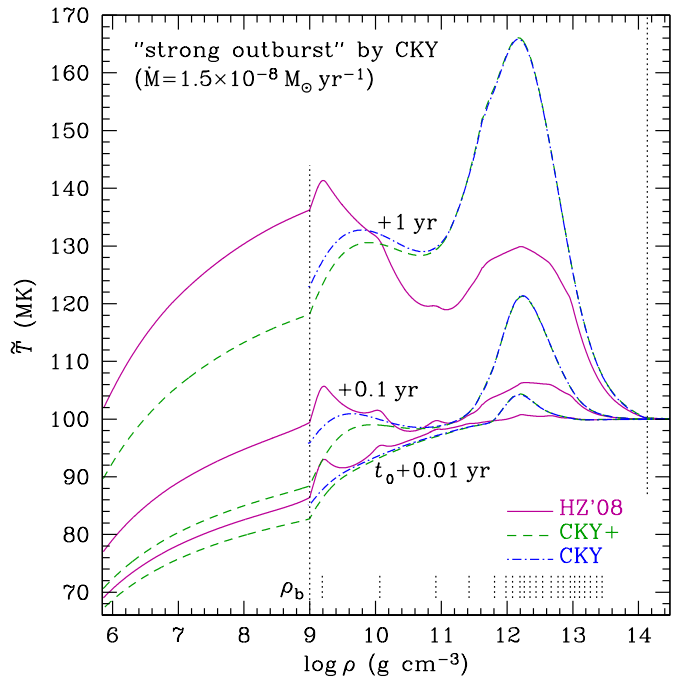}
\caption{Redshifted temperature as function of mass density at selected
time moments (marked near the curves) during
the outburst for the same three models of the crust as in
Fig.~\ref{fig:tcky}.  The labels in the left mark the time since
the start of the  outburst. The redshifted temperature in the core is
$\tilde{T}=10^8$~K. The long vertical dotted lines mark the
boundaries of the crust with the envelope and the core. The short 
vertical dotted lines mark the positions of the heat
sources in the HZ'08 model.
}
\label{fig:rckyH}
\end{figure}
%%%%%%%%%%%%%%%%%%%%%%%%%%%%%%%%%%%%%%%%%%%%%%%%%%%%%%%%%

%%%%%%%%%%%%%%%%%%%%%%%%%%%%%%%%%%%%%%%%%%%%%%%%%%%%%%%%%
\begin{figure}
\centering
\includegraphics[width=\columnwidth]{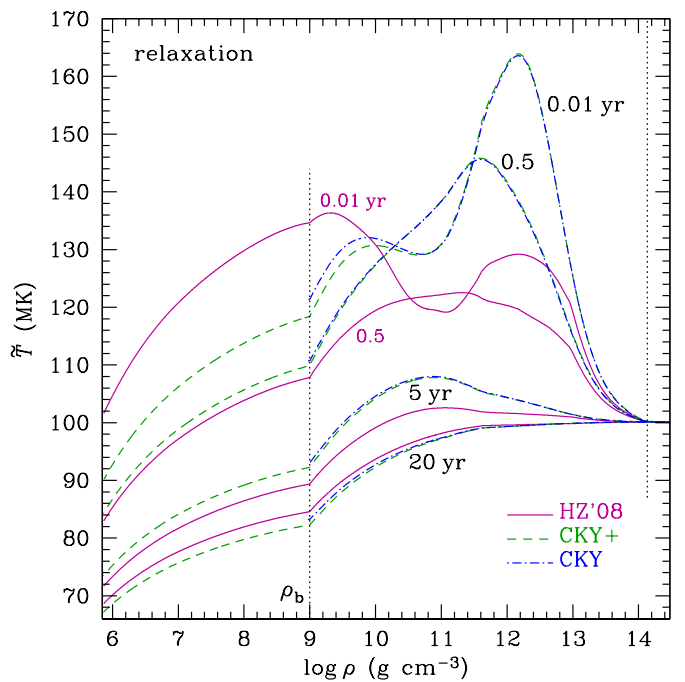}
\caption{The same as in Fig.~\ref{fig:rckyH} but in the quiescent
state.  The labels mark the time since the end of the outburst.
}
\label{fig:rckyR}
\end{figure}
%%%%%%%%%%%%%%%%%%%%%%%%%%%%%%%%%%%%%%%%%%%%%%%%%%%%%%%%%

Figure~\ref{fig:tcky} shows the effective temperature (in energy units)
as a function of time for
a ``strong outburst'' model of \citet{CKY}. The mass accretion rate is
$\dot{M}=1.5\times10^{-8}\,M_\odot$ (slightly below the Eddington limit)
during 1 year. Before the outburst, the star has a quasi-equilibrium
temperature distribution, which corresponds to $\tilde{T}=10^8$~K at the
core/crust interface. Here, $\tilde{T}=\erm{\Phi}T$, where $T$ is the
temperature in the local reference frame and $\Phi$ is the dimensionless
metric function in the Schwarzschild coordinates
\citep[e.g.,][Chapter~32]{MisnerThorneWheeler}. The heat flux $F$ toward
the stellar surface is converted into the effective temperature
according to $F=\sigma_\mathrm{SB}\Teff^4$, where $\sigma_\mathrm{SB}$
is the Stefan-Boltzmann constant, and redshifted to
$\tTeff=\Teff/(1+\zg)$, where $\zg=(1-\rg/R)^{-1/2}-1$ is the
gravitational redshift at the surface, $\rg=2GM/c^2$ is the
Schwarzschild radius, $G$ is the gravitational constant, and $c$ is the
speed of light. The simulations have been performed using the numerical
code described in \citet{PC18}. The BSk24 model of the EoS is adopted
\citep{Pearson_18}. As in the CKY work, the critical temperature for
neutron superfluidity in the crust as function of density is evaluated
using the GIPSF parametrization of \citet{Ho_15}, based on a theoretical
model computed by \citet{Gandolfi_09}. 

The CKY calculations assume that the heat transport through the outer
layers at $\rho<10^9$ \gcc{} is sufficiently quick, so that the thermal
structure of these layers can be treated as quasi-stationary. To test
the effect of this assumption, we compare the CKY model (the dot-dashed
curve in Fig.~\ref{fig:tcky})  with the CKY+ model (the dashed curve),
where the envelope at $\rho<10^9$ \gcc{} is treated more accurately,
uniformly with the internal region. The comparison shows that the
maximum at $t\sim1-2$ yr is less pronounced. A more important difference
is the slower decrease of the luminosity at early cooling before this
maximum. Still greater difference is seen between the lightcurves
calculated using the smooth distribution of the heat sources in the
crust (models CKY and CKY+) and the one where the sources are located at
a series of spherical shells (HZ'08). In the HZ'08 model, the effective
temperature reaches a substantially higher value  at the end of the
outburst, the maximum on the relaxation hillside dissolves, and the
decrease of luminosity becomes monotonic.

The origins of these differences between the lightcurves can be
recognized by considering the thermal structure of the envelope and the
crust during the outburst and relaxation, which is shown in
Figs.~\ref{fig:rckyH}, \ref{fig:rckyR}, and comparing with the heat source distributions
(Fig.~\ref{fig:heatsrc}). Localized heating at densities
$\rho\sim10^9-10^{10}$ \gcc{} cause a quicker rise of temperature at
these densities than the smooth distribution of the heat sources.  In
contrast, the most powerful heat sources localized around
$\rho\sim10^{12}-10^{13}$ \gcc{} raise the temperature less strongly
than the distributed heating source with nearly the same integral heat
release. 

The weaker heating of the outer layers by the CKY and CKY+ models is
easy to explain by looking at the solid and dashed lines in the inset of
Fig.~\ref{fig:heatsrc}. Since the heating is absent at $\rho<\rhob$, the
smooth distribution, being integrated from $\rhob$ inwards, is unable to
provide the same total heat as the localized source.

The stronger temperature increase at $\rho\sim10^{12}$ \gcc{} can also
be explained by comparison of the smooth and localized heat source
distributions in Fig.~\ref{fig:heatsrc}. Although both distributions
provide nearly the same total heat release at
$\rho\lesssim5\times10^{12}$ \gcc, the smooth distribution provides
more energy by $\rho\lesssim3\times10^{12}$ \gcc. Since
the integrated heat capacity is smaller for the thinner outer part of
the crust, such redistribution of the heating causes the stronger
increase of temperature.

%%%%%%%%%%%%%%%%%%%%%%%%%%%%%%%%%%%%%%%%%%%%%%%%%%%%%%%%%
\begin{figure}
\centering
\includegraphics[width=\columnwidth]{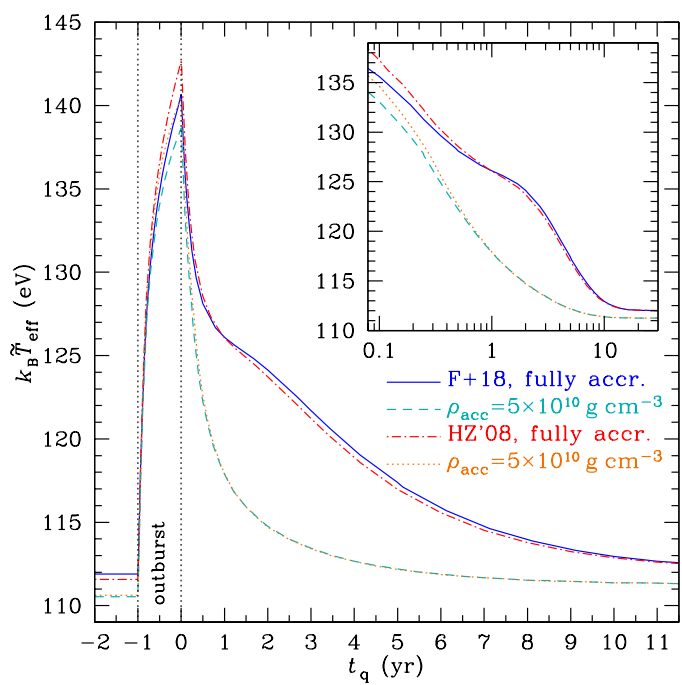}
\caption{Lightcurves for the same neutron star model and outburst
model  as in Fig.~\ref{fig:tcky}, computed for either fully accreted
crust (solid and dot-dashed lines) or partially accreted crust with
replaced matter filling the layer up to
$\rho_\mathrm{acc}=5\times10^{10}$ \gcc{} (dashed and dotted lines) for
models HZ'08 (solid and dashed lines) and F+18 (dot-dashed and dotted
lines), as shown in the legend.
}
\label{fig:t8vs18}
\end{figure}
%%%%%%%%%%%%%%%%%%%%%%%%%%%%%%%%%%%%%%%%%%%%%%%%%%%%%%%%%

%%%%%%%%%%%%%%%%%%%%%%%%%%%%%%%%%%%%%%%%%%%%%%%%%%%%%%%%%
\begin{figure}
\centering
\includegraphics[width=\columnwidth]{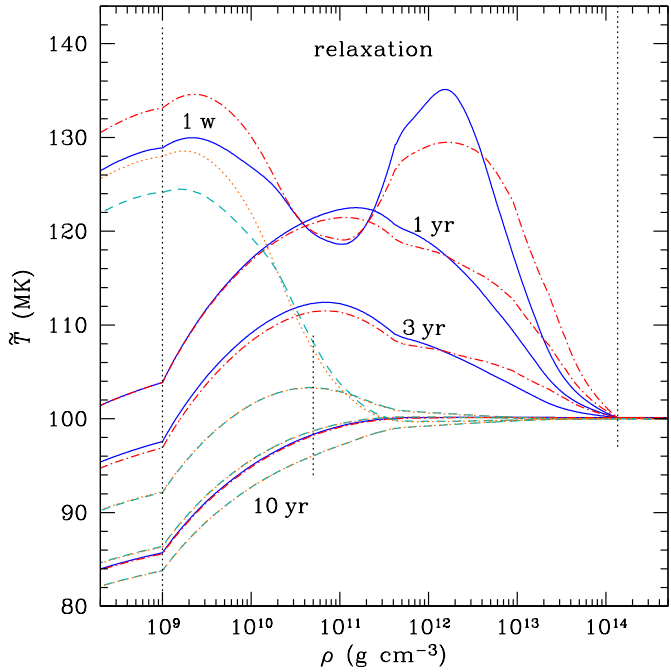}
\caption{Redshifted temperature as function of mass density
at 1 week and 1, 3, and 10 years
after the outburst
for the same models of the crust as in Fig.~\ref{fig:t8vs18}.
The redshifted temperature in the core is fixed at $\tilde{T}=10^8$~K.
The long vertical dotted lines mark the boundaries of the crust
with the envelope and the core. The shorter vertical dotted line marks
the interface between the accreted and ground-state matter
in the models of partially accreted
envelope.
}
\label{fig:r8vs18}
\end{figure}
%%%%%%%%%%%%%%%%%%%%%%%%%%%%%%%%%%%%%%%%%%%%%%%%%%%%%%%%%

In Fig.~\ref{fig:t8vs18} we compare the lightcurves computed with the
HZ'08 and F+18 models. It turns out that the 25\% difference in the
total heat power has a minor effect on the surface emission, because
this difference pertains to large densities (cf.{}
Fig.~\ref{fig:heatsrc}), where the heat leaks mostly to the core. The
temperature distributions at four moments during the crust cooling are
shown in Fig.~\ref{fig:r8vs18}.

The results shown in Fig.~\ref{fig:t8vs18} and in the following figures
are obtained using the MSH model \citep{MargueronSH08} for the neutron
singlet-type superfluidity gap in the crust, as parametrized by
\citet{Ho_15}. We obtain almost identical results with the GIPSF
parametrization employed in Figs.~\ref{fig:tcky} -- \ref{fig:rckyR}.
Results of the more recent extensive numerical simulations of the
neutron singlet superfluidity by \citet{Ding_16} are also very close to
the MSH model.

An amount of accreted matter may be insufficient to fill the entire
crust \citep[e.g.,][]{WijnandsDP13,Fantina_18}, if the total mass of
accreted matter is less than $\sim0.01\,M_\odot$ (cf.{}
Fig.~\ref{fig:heatsrc}). An incomplete replacement of the crust affects
the long-term thermal evolution (see Paper~I) as well as the
post-outburst crustal cooling \citep{CKY}. For illustration, in
Figs.~\ref{fig:t8vs18} and \ref{fig:r8vs18} we compare the lightcurves
and internal temperature distributions in the cases where the accreted
matter fills the entire crust or only the layer at
$\rho<\rho_\mathrm{acc}=5\times10^{10}$ \gcc. For the chosen neutron
star model, the latter case implies $\sim5\times10^{51}$ accreted
baryons.  Assuming the same microphysics as in Paper~I, this
amount of accreted material is consistent with the required
$\tilde{T}=10^8$~K in the core, provided that the accretion lasts 41.6
kyr at the average rate $\langle\dot{M}\rangle=10^{-10}\,M_\odot$
yr$^{-1}$. The heat that is stored only in the outer layers escapes
quicker to the surface, while the heat that is stored deeper in the
crust needs more time to escape. This is reflected in the different
slopes and shapes of the crust cooling lightcurves in
Fig.~\ref{fig:t8vs18}.

%%%%%%%%%%%%%%%%%%%%%%%%%%%%%%%%%%
\section{The effects of ion mixtures in the inner crust}
\label{sect:imp}

%%%%%%%%%%%%%%%%%%%%%%%%%%%%%%%%%%
\subsection{The effect of crust impurities on conductivities}

Post-outburst relaxation of the quasi-persistent SXTs is sensitive to
the crust structure and composition. For example, \citet{Shternin_07}
demonstrated that the observed relaxation of the quasi-persistent SXT KS
1731$-$260 can be reproduced by the theory only if the conductivity is
sufficiently high throughout the major part of the crust, which implies
a nearly pure crystalline structure. On the other hand,
\citet{Deibel_17} found that somewhat better fit to the observed
afterburst cooling can be obtained, if there is a layer with a
significantly lower thermal conductivity near the boundary between
the crust and the core. This is in accord with the findings by
\citet{PonsVR13}, who pointed out that the lack of X-ray pulsars with
spin periods longer than 12~s may be explained by the presence of a
highly resistive layer in the innermost part of the crust of neutron
stars. The discoveries of longer periods of 16.8~s \citep{Hambaryan_17}
and 23.5~s \citep{Tan_18} do not invalidate the qualitative conclusions
of \citet{PonsVR13}. 

The possible existence of a deep layer with low electrical and thermal
conductivities is often attributed to so called ``pasta phase'' of
nuclear matter, which may emerge at the high density, where cylindrical
or plate-like nuclei may constitute the ground state, as predicted by
some models of nuclear matter (e.g.,
\citealp{PethickRavenhall95,Pearson_20}, and references therein). It has
been suggested that the pasta phase may have a low conductivity due to
the  irregularities and spiral defects that are observed in molecular
dynamics (MD) simulations ({e.g., \citealt{Horowitz_15};}
see \citealt{CaplanHorowitz17} for review
and references). These classical MD simulations at temperatures
$T\gtrsim10^{10}$~K and proton fraction $Y_p\sim0.3$\,--\,0.4 might be,
however, more appropriate to studies of the  proto-neutron stars formed
during a supernova burst, than the relaxed neutron stars that have
typically two orders of magnitude lower temperatures and one order of
magnitude smaller proton fraction in the bottom layers of the crust. At
a high temperature of a proto-neutron star, thermal fluctuations should
cause deviations from the regular structure and accordingly
suppression of conductivities not only for the pasta phases but also for
the spherical nuclei. Indeed, MD simulations by \citet{NandiSchramm18}
suggest that the structure factors and hence conductivities in the
pasta phase do not substantially differ from those in the ordinary phase
of quasi-spherical nuclei at the same temperature.

The structure and composition of the inner crust of a mature neutron
star is usually studied in the zero-temperature approximation (see,
e.g., \citealt{HPY}). Then the nuclei of any shape (spherical or not)
are arranged in regular structures. There is no reason for the
conductivity to be lower in the structures composed of rods or slabs
than in the crystal of spherical nuclei. However, the conductivity can
be suppressed by electron scattering off defects or impurities
\citep{Ziman}. 

We will restrict ourselves by the ordinary
phase of quasi-spherical nuclei and calculate the electrical and thermal
conductivities using the code at
\url{http://www.ioffe.ru/astro/conduct/} (see \citealt{PPP15} for review
and references behind this  code\footnote{In Appendix~A.4 of the journal
version of \citet{PPP15}, the words ``Appendix A.3''  (a typographic
error) should read ``equation (A.3)'' in all instances.}). It is
convenient to express electrical conductivity $\sigma$  and thermal
conductivity $\kappa$ in strongly degenerate electron-ion plasmas in
terms of the effective frequencies $\nu_\sigma$, $\nu_\kappa$ of
electron collisions \citep[e.g.,][]{Ziman,YakovlevUrpin80}
\beq
 \sigma = \frac{\nel e^2}{\mel^\ast \nu_\sigma},
\quad \kappa = \frac{\pi^2}{3}\frac{\kB^2 T\nel}{\mel^\ast\nu_\kappa},
\label{elementary} 
\eeq
 where $\nel$ is the electron number density, $e$ is the elementary
charge, $\kB$ is the Boltzmann constant,  $\mel^\ast =
\epsilon_\mathrm{F}/c^2$, and $\epsilon_\mathrm{F}$ is the electron
Fermi energy, including the rest energy. The collision frequencies
in the degenerate matter
 can
be approximately reduced to sums of partial frequencies associated with
relevant electron scattering mechanisms, for example \beq
\nu_{\sigma,\kappa} = \nu_{\sigma,\kappa}^{(\text{ei})} +
\nu_\text{imp}, \label{nu_sum} \eeq where
$\nu_{\sigma,\kappa}^{(\text{ei})}$ and $\nu_\text{imp}$ are the
effective electron-ion and electron-impurity scattering frequencies,
respectively.

The electron-ion collision
frequency in a Coulomb liquid can be written in the form
\citep[e.g.,][]{YakovlevUrpin80}
\begin{equation}
   \nu_{\sigma,\kappa}^{(\text{ei})} =
   4\pi Z^2 e^4 \nion\,\mel^\ast\, p_\mathrm{F}^{-3}
    \Lambda_{\sigma,\kappa}(p_\mathrm{F}),
\label{tau}
\end{equation}
where $\nion$ is the ion number density, $p_\mathrm{F}$ is the electron
Fermi momentum, and $\Lambda$ is a dimensionless Coulomb logarithm.
\citet{Potekhin_99} derived a unified treatment of the
conductivities due to degenerate electrons
in the Coulomb liquid and Coulomb crystal and described
both regimes by \req{tau}. In this formalism, by order of magnitude,
$\Lambda\sim1$ in the ion liquid, and $\Lambda\sim T/T_\mathrm{m}$ in
the pure Coulomb crystal with a melting temperature $T_\mathrm{m}$.

The mean frequency of electron scattering on impurities with different
charge numbers $Z_j$ is described, following \citet{YakovlevUrpin80},
using the substitution of $Z^2$ in \req{tau} by the \emph{impurity
parameter}
\beq
Q_\text{imp} = \sum_j Y_j (Z_j - \langle Z \rangle )^2,
\label{Qimp}
\eeq
that is
$
   \nu_\text{imp} = 
   4\pi Q_\text{imp} e^4 \nion\,\mel^\ast\, p_\mathrm{F}^{-3}
    \Lambda_\text{imp}.
$
Here, $Y_j$ is the number fraction of the nuclei of the $j$th kind, and
$\langle Z \rangle \equiv \sum_j Y_j Z_j$ is the mean charge number.  It
is important that the impurities are randomly distributed. Therefore,
the electron-impurity collision frequencies are not suppressed by the
crystal long-ordering, and the Coulomb logarithm $\Lambda_\text{imp}$
does not include the  factors that suppress the scattering rate in a
crystal \citep[see][]{Potekhin_99}. For this reason, at low temperatures
or high densities the electron-impurity scattering dominates and
controls the conductivities (cf.{} \citealt{GYP01}).

%%%%%%%%%%%%%%%%%%%%%%%%%%%%%%%%%%%%%%%%%%%%%%%%%%%%%%%%%
\begin{figure}
\centering
\includegraphics[width=\columnwidth]{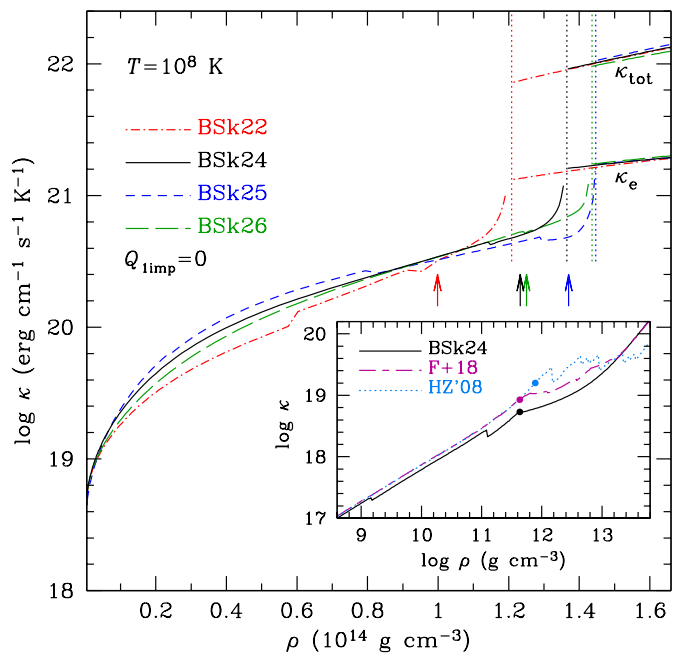}
\caption{Thermal conductivity $\kappa$ 
as a function of mass density $\rho$ in the
inner crust and outer core of a neutron star
for the nuclear energy density functionals 
BSk22, BSk24, BSk25, and BSk26.
Vertical arrows mark the proton drip
densities. Vertical dotted lines mark the crust-core interface.
For the core  densities, the electron (lower lines) and total (electron
and baryon, upper lines) conductivities are shown.
\emph{Inset:} $\kappa$ vs.{} $\rho$ in the outer and inner
crust for the model BSk24 (the ground state) compared with the accreted
crust models F+18 and HZ'08. The dots on the curves mark the neutron
drip for each model.
}
\label{fig:kapt8}
\end{figure}
%%%%%%%%%%%%%%%%%%%%%%%%%%%%%%%%%%%%%%%%%%%%%%%%%%%%%%%%%

For thermal conductivity, electron-electron effective scattering
frequency $\nu_{\kappa}^{(\text{ee})}$ should be added to the sum
(\ref{nu_sum}). We treat it following \citet{ShterninYakovlev06}. The
electron-electron scattering becomes particularly important in the deep
layers of the inner crust of sufficiently cold neutron stars (for
example, in the absence of impurities it dominates at
$\rho\gtrsim3\times10^{13}$ \gcc, if $T\lesssim10^7$~K).

Figure \ref{fig:kapt8} shows the dependence of thermal conductivity
$\kappa$ on mass density $\rho$ in a neutron star without impurities.
The main frame demonstrates the conductivity in the ground state,
computed for the four BSk energy density functionals according to
\citet{Pearson_18}. The electron thermal conductivity in the crust is
computed as described above. The electron thermal conductivity in the
core is computed according to \citet{ShterninYakovlev07}, and the baryon
thermal conductivity in the core is computed according to
\citet{BaikoHY01}. The inset compares the electron thermal
conductivities for the nonaccreted crust and for two models of the fully
accreted crust. The differences for different models are perceptible,
but not dramatic. An important feature is the conductivity increase with
density, which accelerates beyond the proton drip because the nuclei
become gradually dissolved in this transitional layer, so that an
approximate continuity of the electron thermal conductivity between the
crust and the core is observed. In the core, however, heat transport by
degenerate neutrons dominates, which increases the thermal conductivity
by almost an order of magnitude. For these reasons, the core of a not
too young neutron star is nearly isothermal.

%%%%%%%%%%%%%%%%%%%%%%%%%%%%%%%%%%
\subsection{Nuclear mixtures in the crust}
\label{sect:impcrust}

A standard assumption concerning the crust composition of a neutron
star  is electrically charge neutral matter in its absolute ground state
\citep{Harrison_65}. The composition of any crustal layer at pressure
$P$ is thus obtained from the absolute minimum of the Gibbs free energy
per nucleon. The actual crust formation in a newly born neutron star
proceeds from an extremely hot ($T\gg10^{10}$~K) initial state in the
aftermath of gravitational collapse \citep[e.g.,][]{KeilJanka95}, 
but in $\sim 1-10$ yr it cools down
to $T\lesssim10^9$~K \citep[e.g.,][]{GYP01,PC18}.

Hot plasma in the outer layers of a proto-neutron star  after
deleptonization is initially (while $T\gtrsim10^{10}$~K) in nuclear
statistical equilibrium (NSE), which is assured by the huge speed of
photo-disintegration reactions, which destroy nuclei, and radiative
captures, which build nuclei \citep[e.g.,][]{Clayton,RolfsRodney,Langer}. 
Assuming beta equilibrium,
the NSE composition is determined not by reaction rates,
but by the relative binding
energies of nuclei, nuclear spins, temperature, and mean baryon density.
Plasma composition is determined by Saha equations involving nuclei,
neutrons, protons, and $\alpha$ particles.
For example,
the Saha equations for nuclei $N(A,Z)$, 
neutrons $N_n$ and protons $N_p$
can be written as
 \citep[see, e.g.,][]{Langer}
\beq
\frac{N(A-1,Z)N_n}{N(A,Z)}=\Theta_n(A,Z,T)
\exp\left(-\frac{Q_n}{\kB T}\right)
\label{Saha_n}
\eeq
and
\beq
\frac{N(A-1,Z-1)N_p}{N(A,Z)}=\Theta_p(A,Z,T)
\exp\left(-\frac{Q_p}{\kB T}\right),
\label{Saha_p}
\eeq
where $Q_n$ ($Q_p$) is a neutron (proton) binding energy 
in the nucleus $(A,Z)$, and a factor $\Theta_n$ ($\Theta_p$) 
is proportional to $T^{2/3}$ and to
the ratio of statistical weights of the considered nuclei.

%%%%%%%%%%%%%%%%%%%%%%%%%%%%%%%%%%%%%%%%%%%%%%%%%%%%%%%%%
\begin{figure}
\centering
\includegraphics[width=.48\columnwidth]{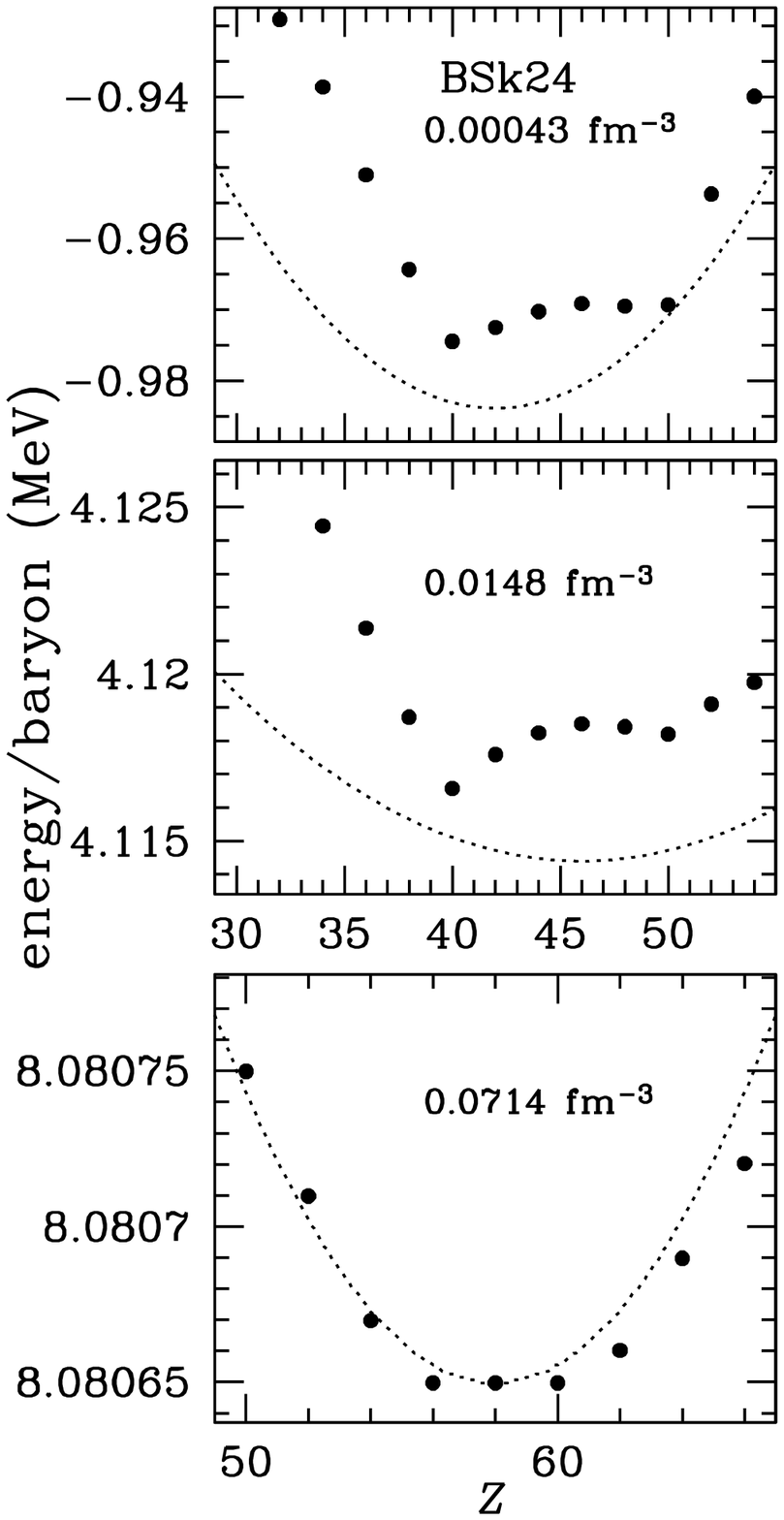}
\includegraphics[width=.49\columnwidth]{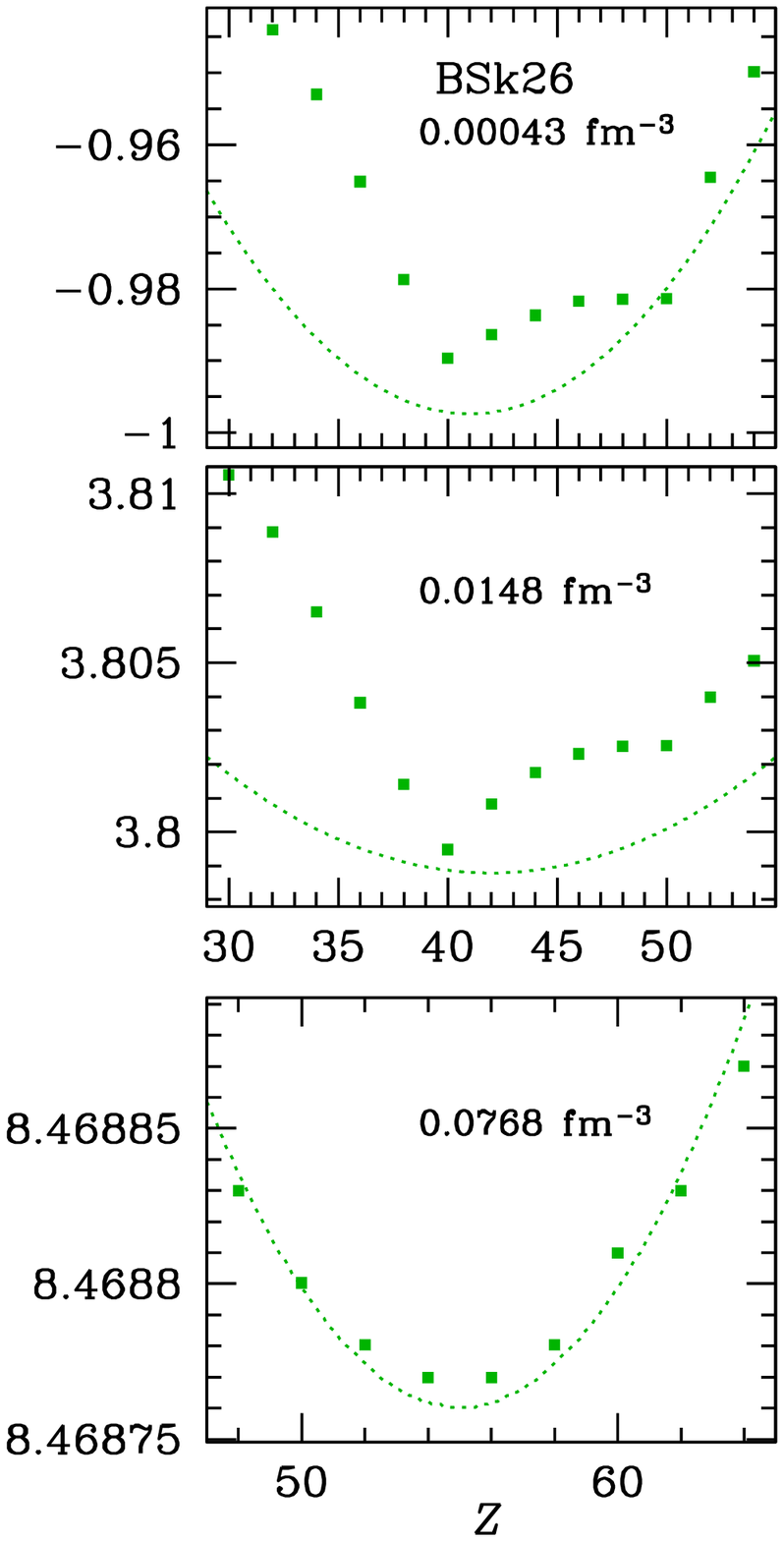}
\caption{Energy per baryon as a function of the nuclear charge number
for the functionals BSk24 (left) and BSk26 (right) at densities near the
top, in the middle, and near the bottom of the inner crust (the top,
middle, and bottom panels, respectively). The symbols show the computed
energy values, and the dotted line shows the parabolic approximation
near the energy minimum in the ETF approximation.
}
\label{fig:energy24}
\end{figure}
%%%%%%%%%%%%%%%%%%%%%%%%%%%%%%%%%%%%%%%%%%%%%%%%%%%%%%%%%

%%%%%%%%%%%%%%%%%%%%%%%%%%%%%%%%%%%%%%%%%%%%%%%%%%%%%%%%%
\begin{figure}
\centering
\includegraphics[width=.9\columnwidth]{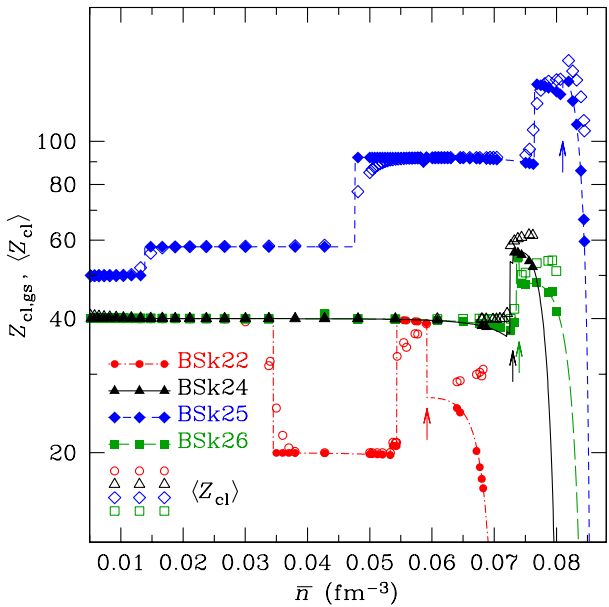}
\caption{Ground state charge number of a nucleon cluster 
$Z_\mathrm{cl,gs}$ (lines and filled symbols) and equilibrium charge
number $\langle Z_\mathrm{cl}\rangle$, at the ``freeze-out'' temperature
$\Ta=10^{9.5}$\,(K) (open symbols), evaluated using the tables
of energies $e(\nbar,Z)$ published by \citet{Pearson_18} for the nuclear
energy density functionals  BSk22, BSk24, BSk25, and BSk26. Vertical
arrows mark the proton drip densities.
}
\label{fig:zmean}
\end{figure}
%%%%%%%%%%%%%%%%%%%%%%%%%%%%%%%%%%%%%%%%%%%%%%%%%%%%%%%%%

Equations (\ref{Saha_n}), (\ref{Saha_p}) show that the rates of
reactions assuring NSE are strongly $T$-dependent. When the matter cools
below some temperature $\Ta\sim\mbox{a few}\times10^9$~K, nuclear
composition ``freezes'' (ceases to change), because relevant reaction
channels (dissociation, absorption) become closed
\citep{Clayton,RolfsRodney,Langer}. Then one can use a rough
approximation of neglecting thermal effects (see, e.g., Fig.~3.1 of
\citealt{HPY}). 

Still the question remains as to how close NSE is, at $T\sim\Ta,$ to the
absolute minimum energy state. Figure~\ref{fig:energy24} shows the
dependence of the  energy per baryon as a function of the nuclear
charge number $Z$, according to \citet{Pearson_18}, for two energy
density functionals BSk24 and BSk26 at three
values of the mean baryon density $\nbar$ in the inner crust, selected near the top, in the
middle, and near the bottom of the inner crust. At the first two
densities, the energies are computed with proton shell and pairing
corrections. The shell corrections provide the sharp minimum at the
``magic number'' $Z=40$. The largest density for each functional is
chosen beyond the ``proton drip'' density $\npd$, where these
corrections largely vanish. In such cases, \citet{Pearson_18} computed
the energy using the extended Thomas-Fermi theory (ETF) without shell
and pairing corrections. The ETF energy changes smoothly with $Z$. Its
approximation by a parabola at the minimum is shown by a dotted line in
each panel of Fig.~\ref{fig:energy24}. The differences between the
energy values at the minimum and at the neighboring values of $Z$ are of
the order of a few keV per baryon at the lower densities, but they fall 
to $\sim0.01$ keV beyond the proton drip. For a Wigner-Seitz cell, which
comprises typically $\sim10^3$ baryons,  the energy differences are of
the order of 1~MeV in the middle of the inner crust and not larger than
tens keV near the crust bottom. In the last case, the energy difference
is smaller than the thermal energy $\kB\Ta$ at the ``freezing'' point of
nuclear composition, so that one should expect to find a mixture of
nuclei with different charge numbers $Z$. 

%%%%%%%%%%%%%%%%%%%%%%%%%%%%%%%%%%%%%%%%%%%%%%%%%%%%%%%%%
\begin{figure}
\centering
\includegraphics[width=.48\columnwidth]{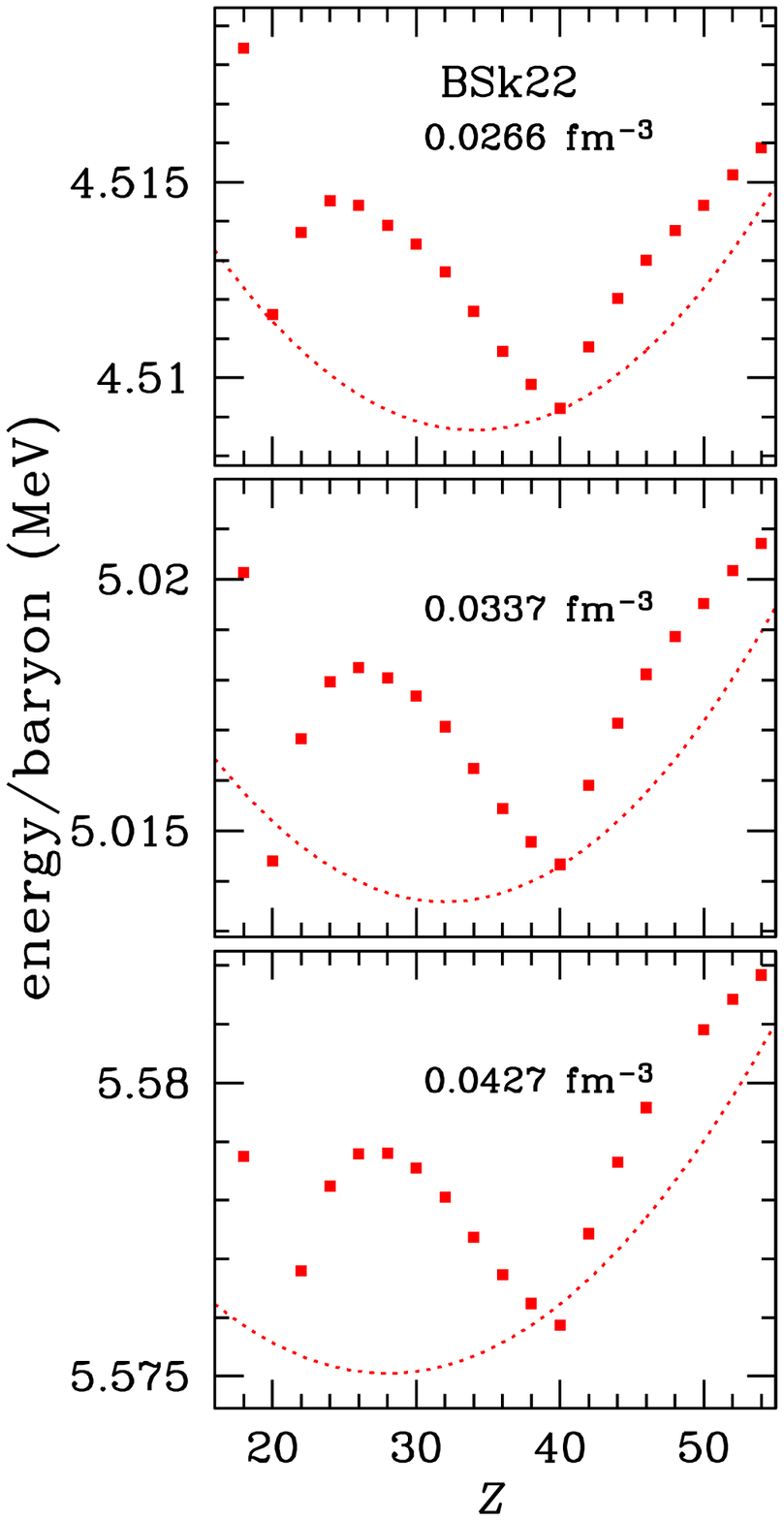}
\includegraphics[width=.49\columnwidth]{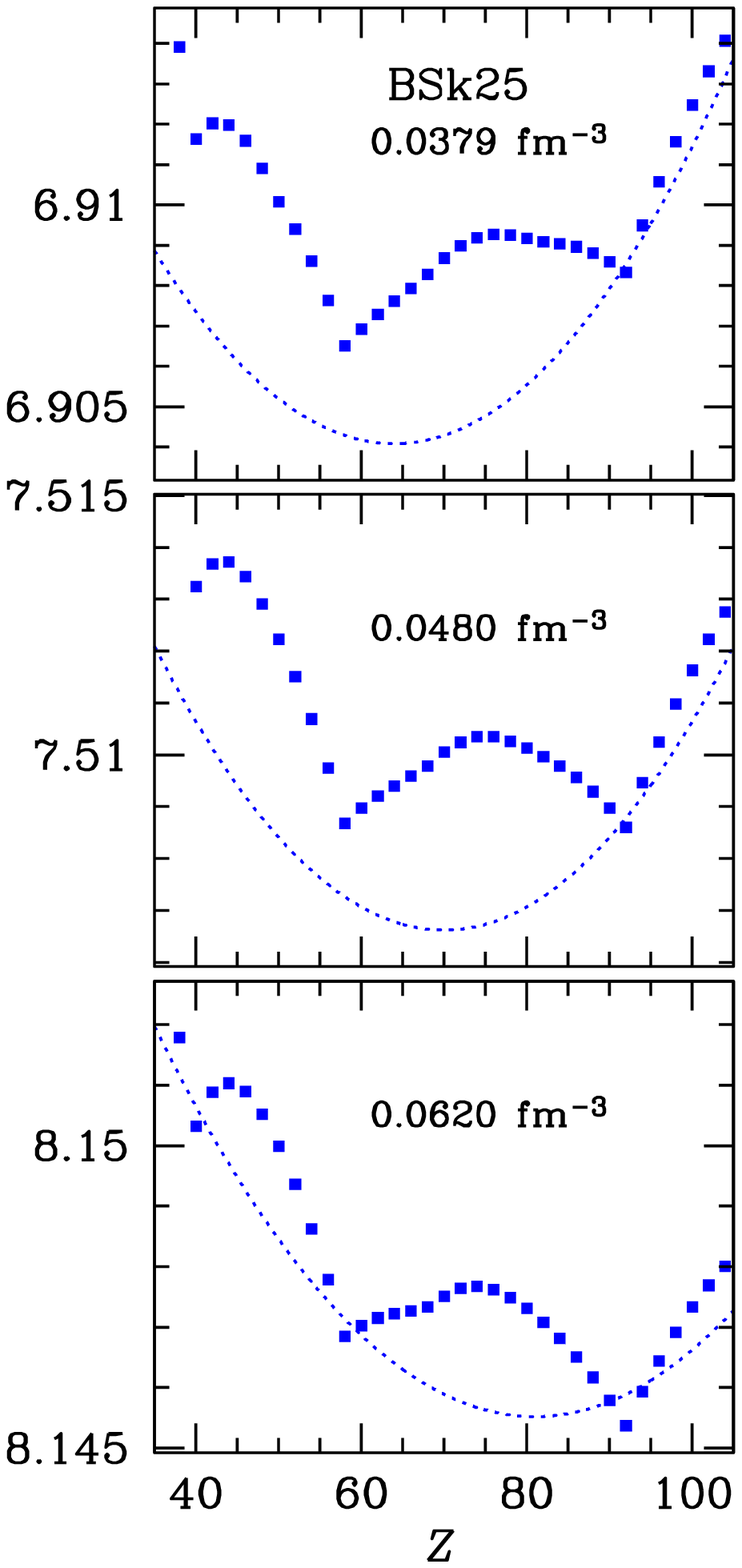}
\caption{The same as in Fig.~\ref{fig:energy24}, but for the models
BSk22 (left panels) and BSk25 (right panels) at three densities $\nbar$
around a jump of $Z_\mathrm{cl,gs}$.
}
\label{fig:energy25}
\end{figure}
%%%%%%%%%%%%%%%%%%%%%%%%%%%%%%%%%%%%%%%%%%%%%%%%%%%%%%%%%

An accurate calculation of the composition could be provided by solving
the equations of kinetic equilibrium, involving all relevant reaction
rates, together with the neutron star cooling equation. This complex
task goes far beyond the scope of our work. Instead, we make a
simple order-of-magnitude estimate of the impurity parameter, based on
the ion sphere approximation and on the Boltzmann statistics in the
vicinity of the ground state.  We treat a mixture of nuclei, free
neutrons, and electrons in the inner crust at a given average baryon
number density $\nbar$ with energies close to the ground state as an
ensemble of ion spheres (i.e., Wigner-Seitz cells in spherical
approximation) with different charge numbers $Z$.  The ion sphere radius
$R_\mathrm{WS}$ is determined by the charge neutrality condition,
$(4\pi/3)\nel R_\mathrm{WS}^3=Ze$. We assume that all the cells are
close to  the absolute ground state. Applying the linear mixing rule for
the dense plasmas, we approximately write the cell energy as
\beq
E(\nbar,Z)=A'(\bar{n},Z)\,e(\bar{n},Z),
\eeq
where $e(\bar{n},Z)$ is energy per baryon in the model of a plasma
composed of a single type of nuclei and $A'$ is the total number of baryons
in the cell. Then small deviations from the ground state, which are
caused by thermal excitations, add the energy
\beq
E_\mathrm{exc}(Z)=A'(\bar{n},Z)\,
     \left[e(\bar{n},Z)-e_\mathrm{gr.st.}\right].
\label{E_exc}
\eeq
Assuming that these excitations obey the Boltzmann statistics, we can
write the statistical weight of the given charge number $Z$ in the
mixture in the form $\Theta(Z,T) \exp(-{E_\mathrm{exc}(Z)}/{\kB T}),$
where $\Theta(Z,T)$ is a factor, which varies much slower than the
exponential at $T\sim \Ta$, so that we neglect this variation. Then
the abundances of different chemical elements are
\beq
Y_Z \approx
\exp\left(-\frac{E_\mathrm{exc}(Z)}{\kB\Ta}\right)
\Big/
\sum_Z
\exp\left(-\frac{E_\mathrm{exc}(Z)}{\kB\Ta}\right).
\label{Y_Z}
\eeq
In numerical examples below, following \citet{Sato79},  we adopt 
$\Ta=10^{9.5}$~K.

%%%%%%%%%%%%%%%%%%%%%%%%%%%%%%%%%%%%%%%%%%%%%%%%%%%%%%%%%
\begin{figure}
\centering
\includegraphics[width=.9\columnwidth]{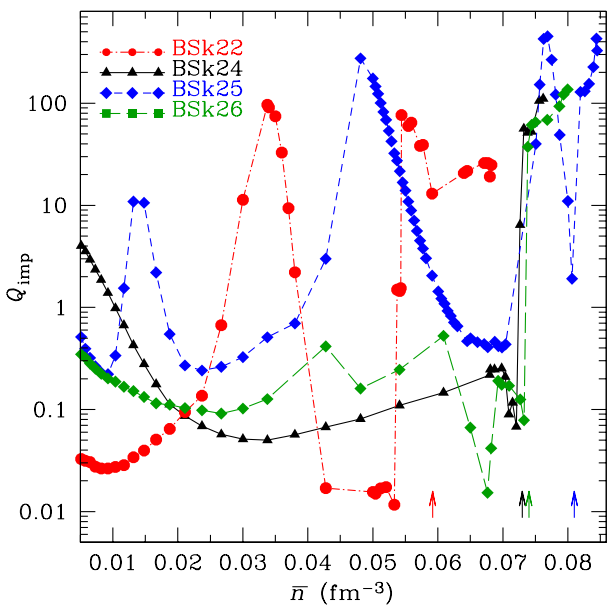}
\caption{Impurity parameter as a function of the mean baryon density for
the nuclear energy density functionals  BSk22, BSk24, BSk25, and BSk26.
The calculated points are connected by lines as guides to eye.
Vertical arrows mark the proton drip densities.
}
\label{fig:qimp}
\end{figure}
%%%%%%%%%%%%%%%%%%%%%%%%%%%%%%%%%%%%%%%%%%%%%%%%%%%%%%%%%

%%%%%%%%%%%%%%%%%%%%%%%%%%%%%%%%%%%%%%%%%%%%%%%%%%%%%%%%%
\begin{figure*}
\centering
\includegraphics[width=.2739\textwidth]{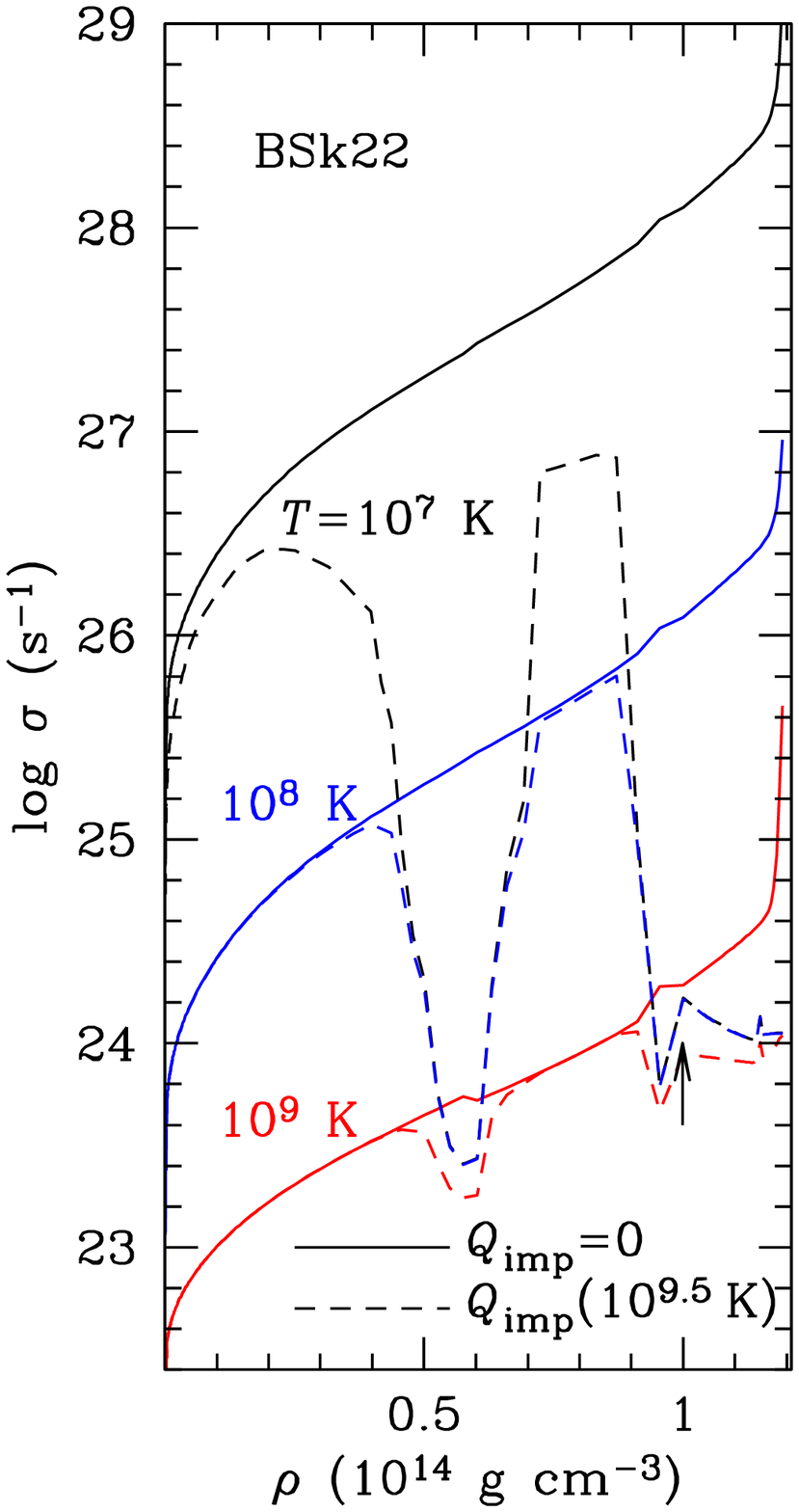}
\includegraphics[width=.22\textwidth]{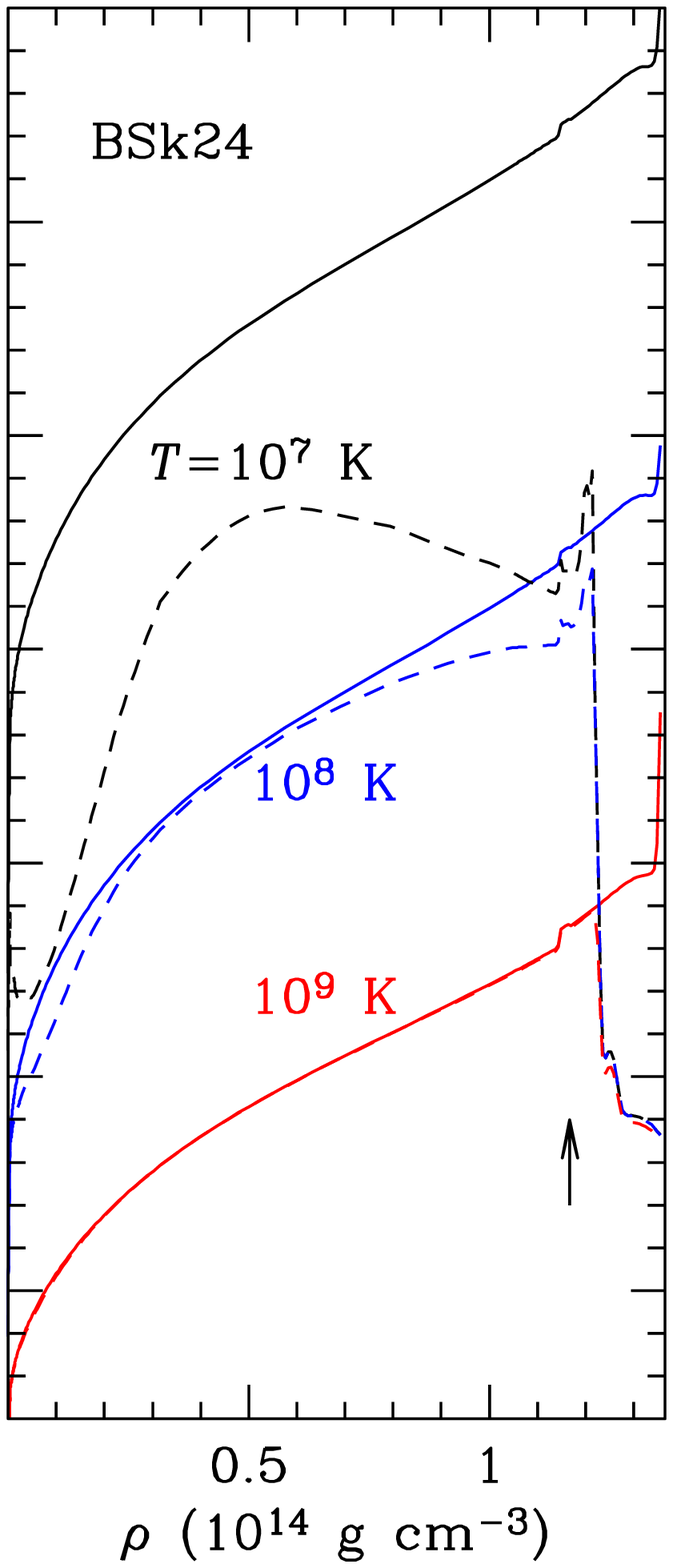}
\includegraphics[width=.22\textwidth]{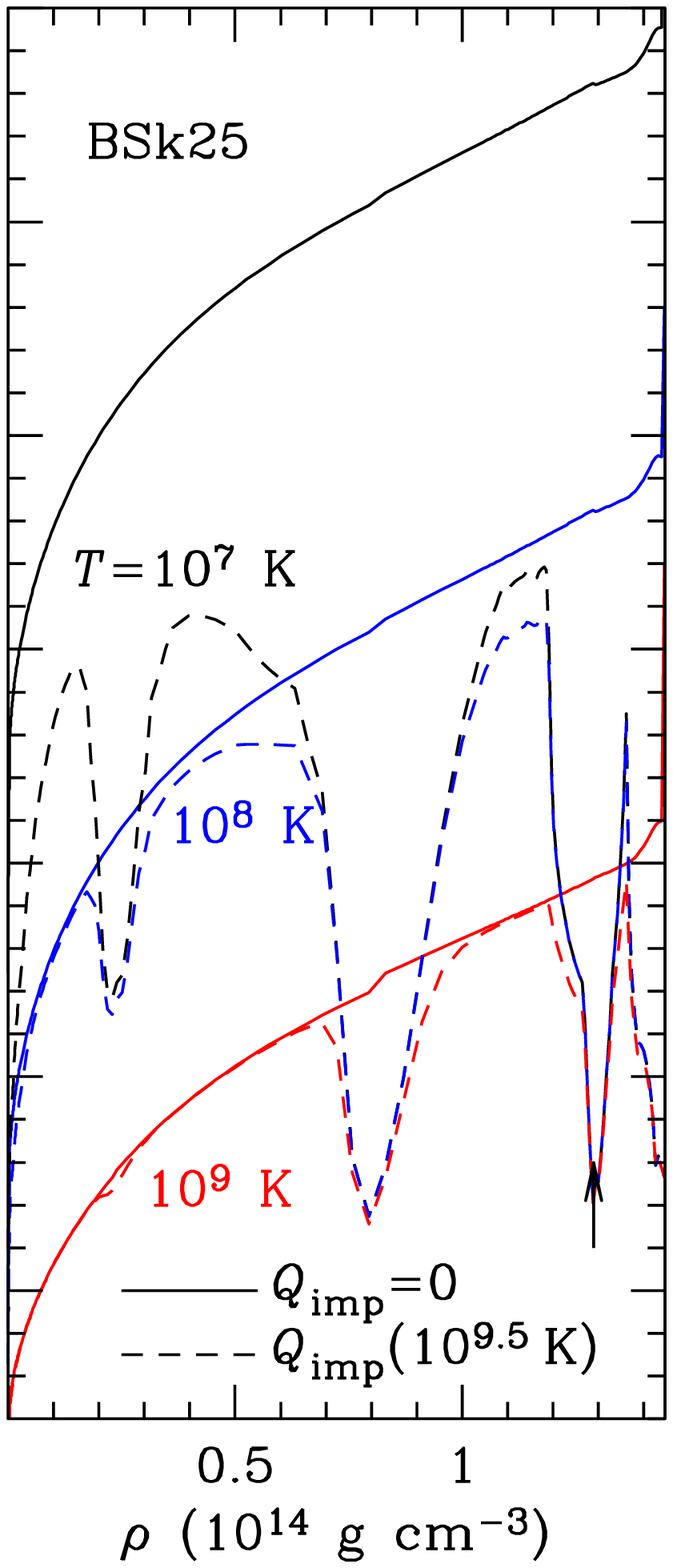}
\includegraphics[width=.22\textwidth]{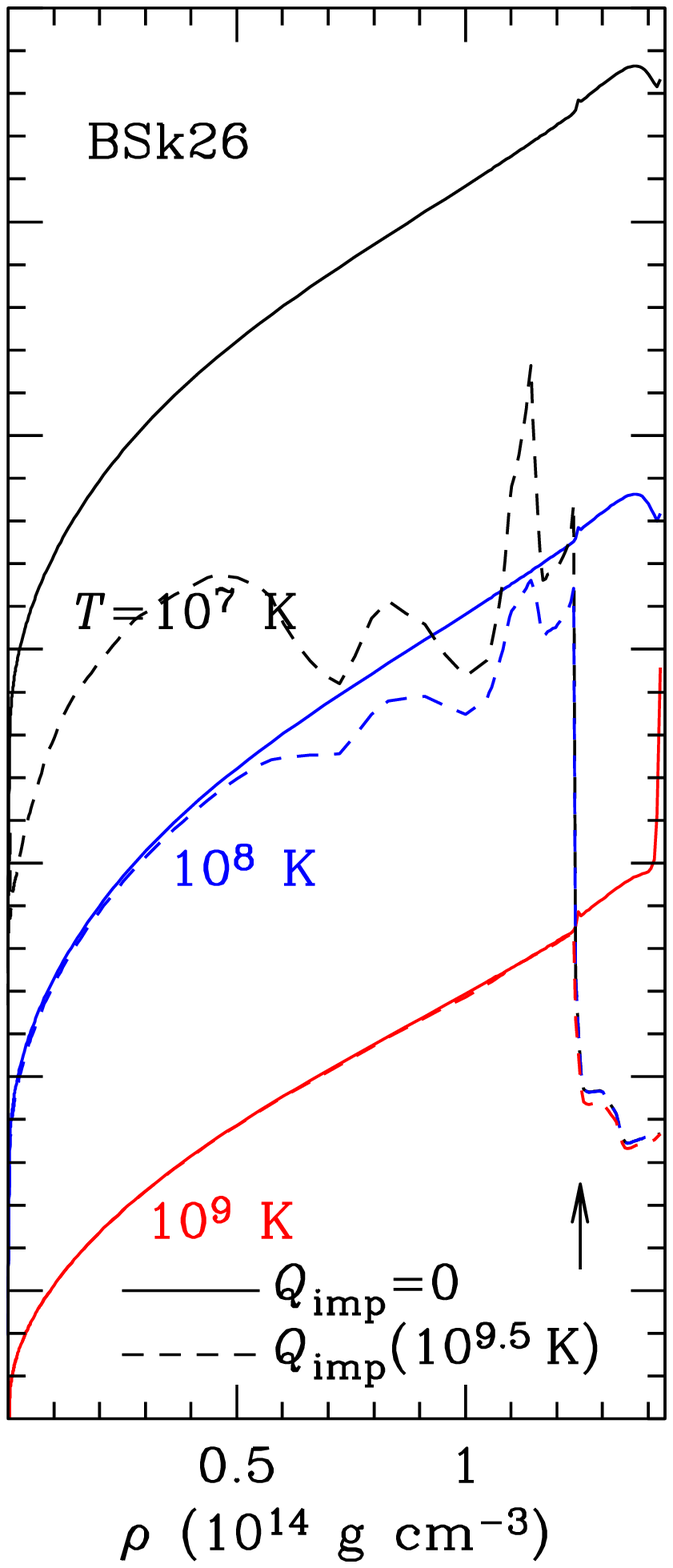}
\caption{Electrical conductivity as a function of the mass density for
the nuclear energy density functionals  BSk22, BSk24, BSk25, and BSk26
(from left to right panels) in a pure inner crust (solid lines) and in
the inner crust with the impurity parameter shown in Fig.~\ref{fig:qimp}
(dashed lines) at temperatures $T=10^7$~K, $10^8$	K, and $10^9$~K.
The arrows mark the proton drip densities.
}
\label{fig:sigma}
\end{figure*}
%%%%%%%%%%%%%%%%%%%%%%%%%%%%%%%%%%%%%%%%%%%%%%%%%%%%%%%%%

%%%%%%%%%%%%%%%%%%%%%%%%%%%%%%%%%%%%%%%%%%%%%%%%%%%%%%%%%
\begin{figure*}
\centering
\includegraphics[width=.48\textwidth]{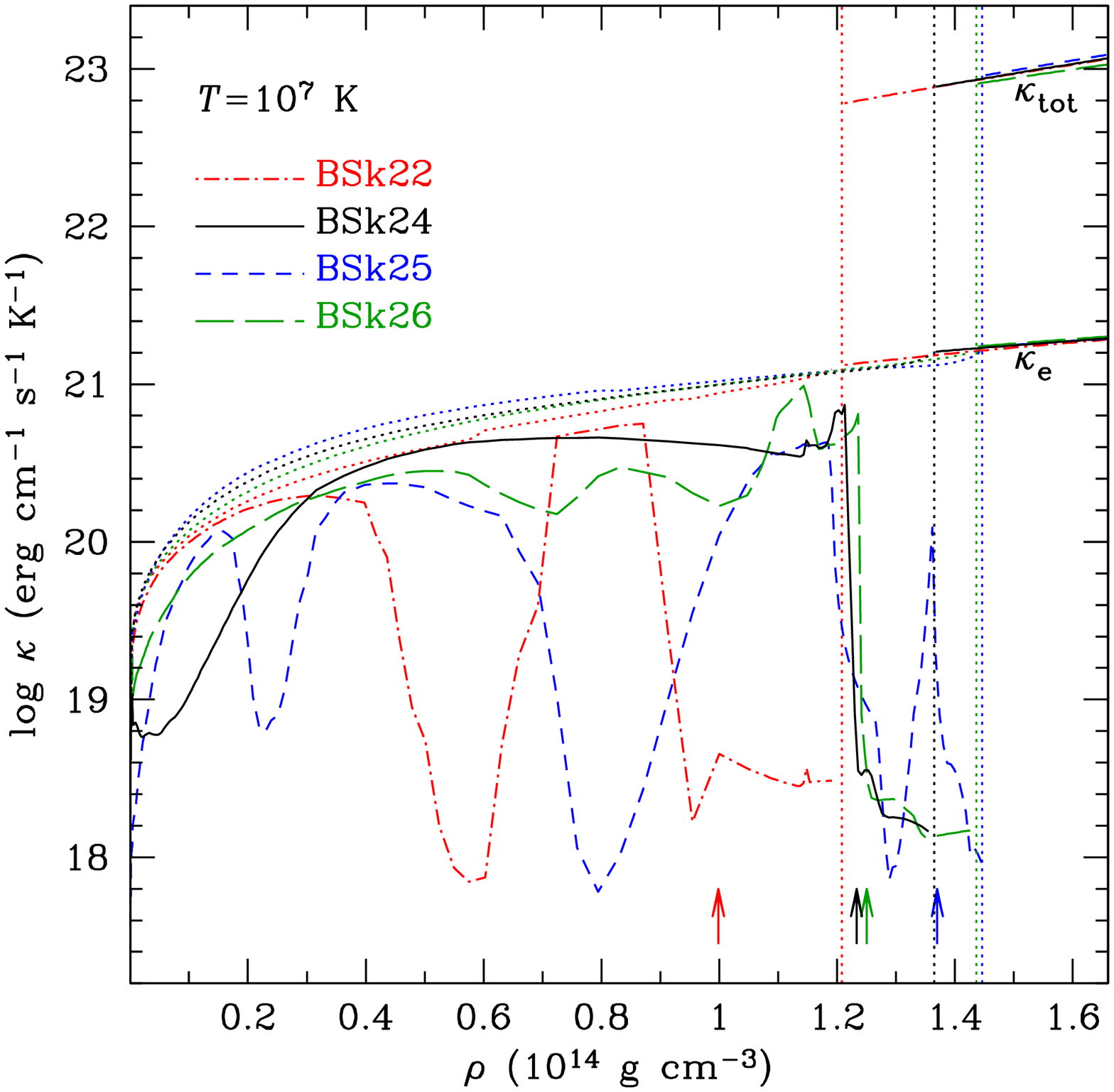}
\includegraphics[width=.475\textwidth]{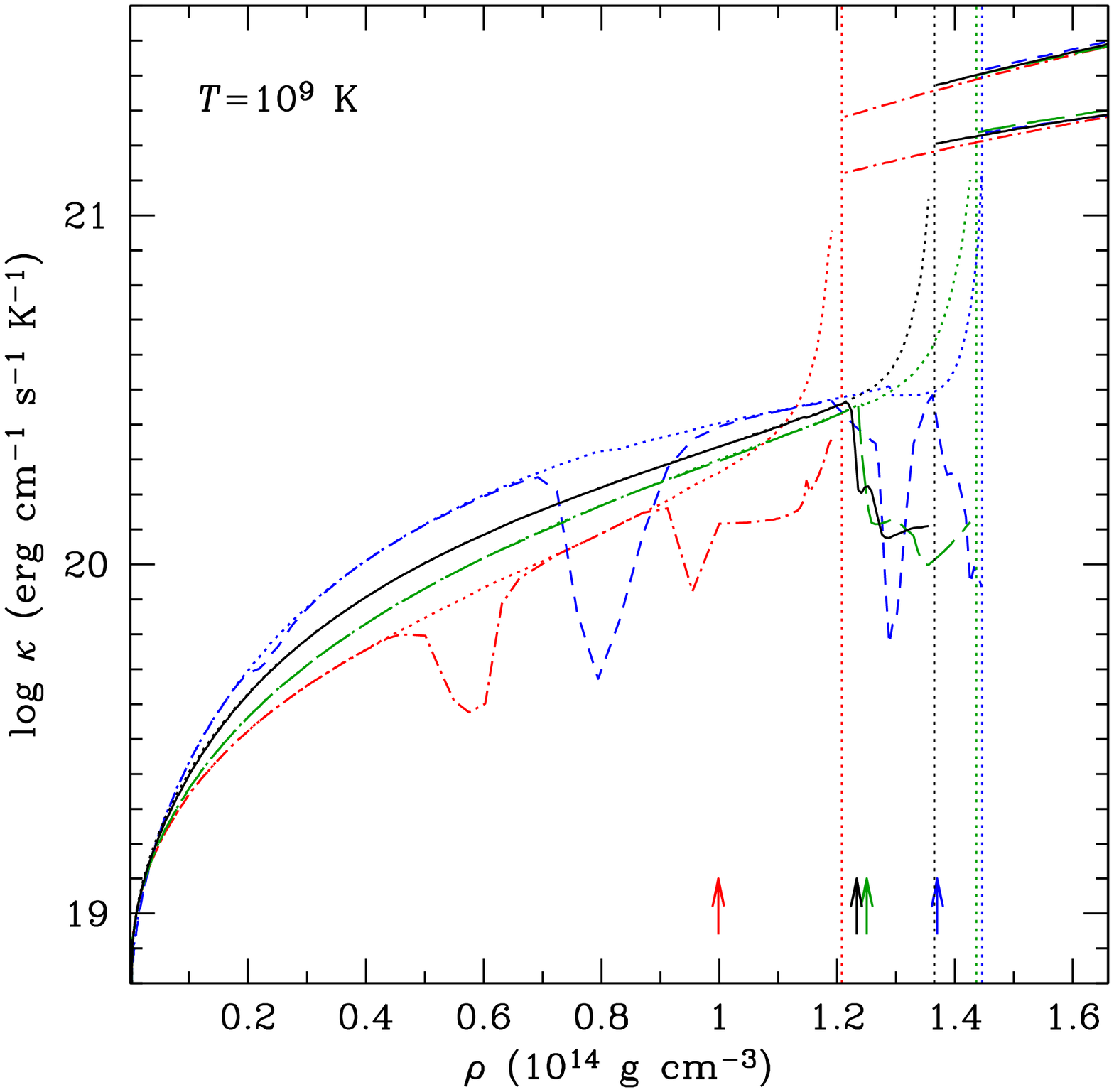}
\caption{Thermal conductivity as a function of the mass density for the
models BSk22, BSk24, BSk25, and BSk26 (dot-dashed, solid, short-dashed
and long-dashed lines, respectively) in the inner crust and in the outer
core. The conductivities are computed with the impurity parameter shown
in Fig.~\ref{fig:qimp} at temperatures $T=10^7$~K (the left panel) and
$10^9$~K (the right panel). For comparison, thermal conductivities in
the pure crust are drawn by the dotted 
curves. As in Fig.~\ref{fig:kapt8}, the arrows mark the proton
drip densities
for the four EoSs, the vertical dotted lines mark the crust-core
interfaces, and to the right of these lines (at the core densities) both
the electron and total conductivities are shown.
}
\label{fig:kappa_imp}
\end{figure*}
%%%%%%%%%%%%%%%%%%%%%%%%%%%%%%%%%%%%%%%%%%%%%%%%%%%%%%%%%

%%%%%%%%%%%%%%%%%%%%%%%%%%%%%%%%%%%%%%%%%%%%%%%%%%%%%%%%%
\begin{figure}[h!]
\centering
\includegraphics[width=\columnwidth]{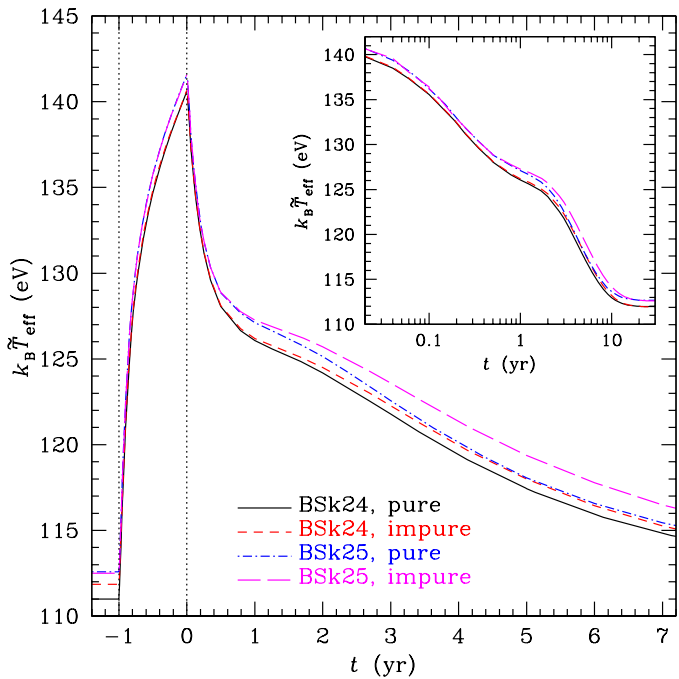}
\caption{Lightcurves for the same outburst model as in
Figs.~\ref{fig:tcky} and \ref{fig:t8vs18}, computed assuming that the
accreted matter fills the crust to the density
$\rho_\mathrm{acc}=10^{13}$ \gcc{} for a neutron star with mass
$M=1.4\,M_\odot$, using the EoS models BSk24 (solid and short-dashed
lines) and BSk25 (dot-dashed and long-dashed lines). The nonaccreted
part of the crust is assumed either pure (solid and dot-dashed lines) 
or with the frozen equilibrium mixture (short- and long-dashed lines).
The F+18 model is adopted for the accreted crust composition and
heating. The inset shows the cooling part of the same lightcurves in
the logarithmic scale.
}
\label{fig:tT100imp}
\end{figure}
%%%%%%%%%%%%%%%%%%%%%%%%%%%%%%%%%%%%%%%%%%%%%%%%%%%%%%%%%

%%%%%%%%%%%%%%%%%%%%%%%%%%%%%%%%%%%%%%%%%%%%%%%%%%%%%%%%%
\begin{figure*}
\centering
\includegraphics[height=.34\textwidth]{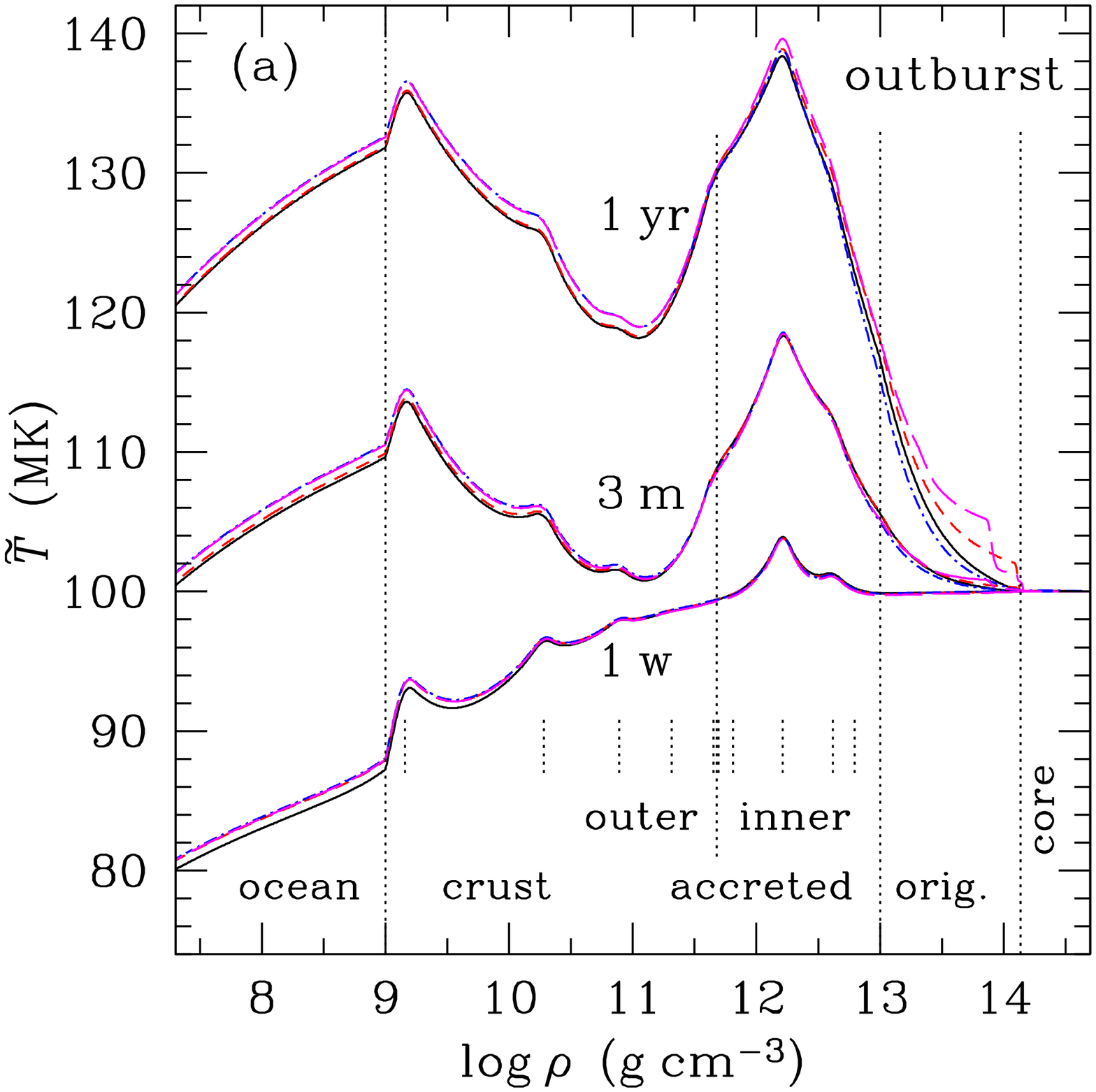}
\includegraphics[height=.34\textwidth]{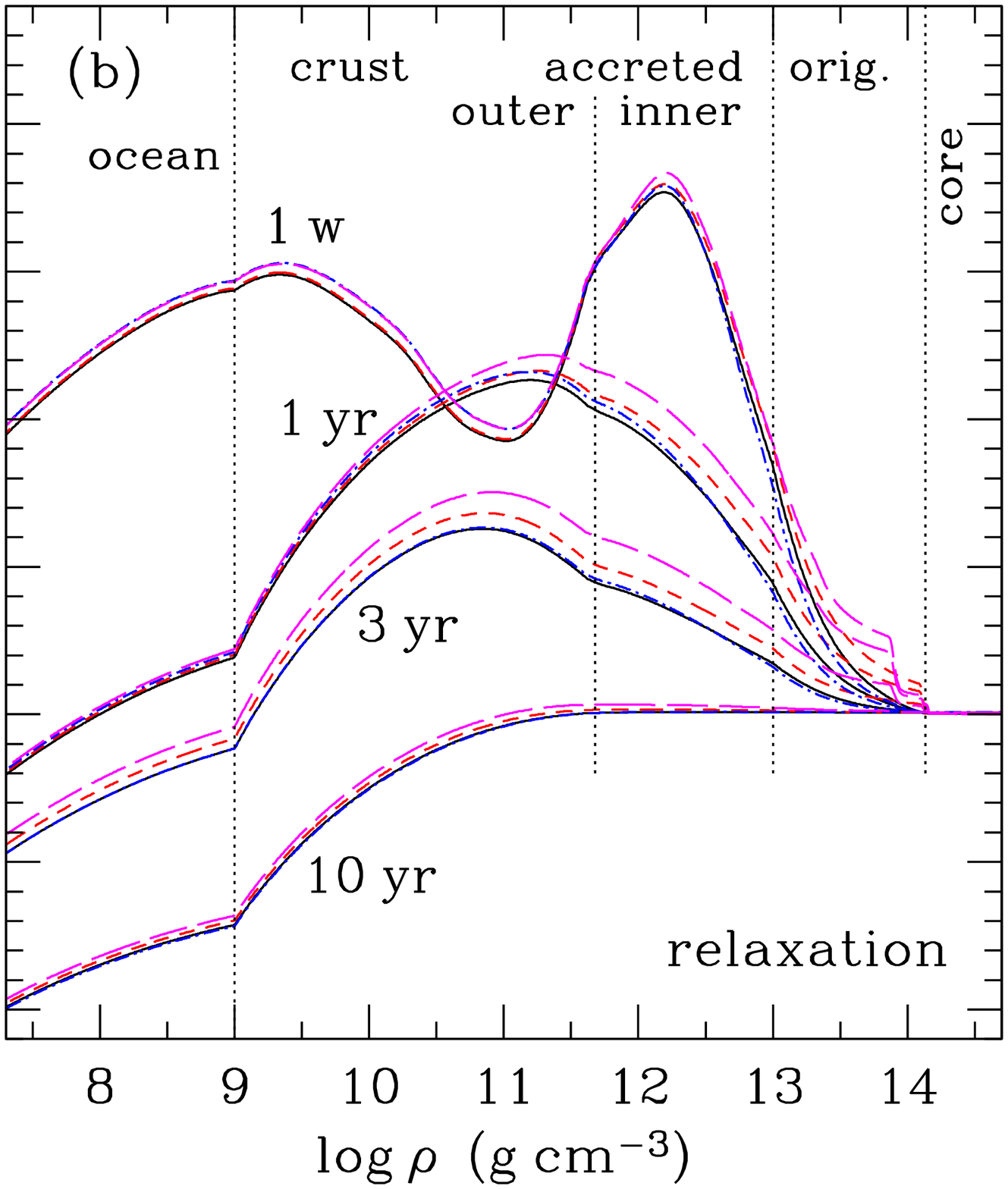}
\includegraphics[height=.34\textwidth]{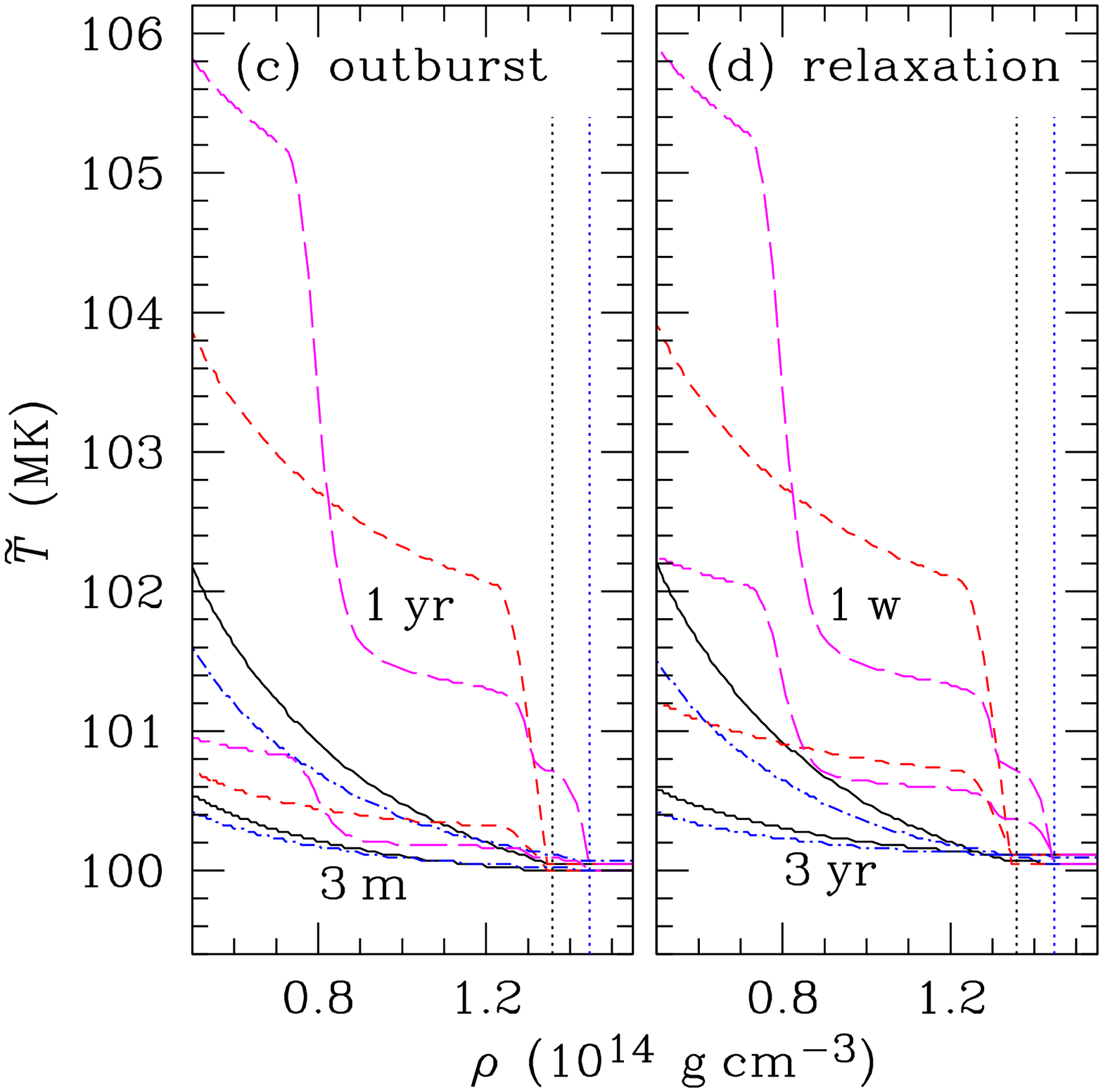}
\caption{Redshifted temperature as function of mass density at selected
time moments (marked near the curves) during  (panel \emph{a}) and after
(panel \emph{b}) the outburst for the same models as in
Fig.~\ref{fig:tT100imp}. The redshifted temperature in the core is fixed
at $\tilde{T}=10^8$~K. The long vertical dotted lines mark the
boundaries between the liquid envelope and the solid crust, the outer
and inner crust, the accreted and nonaccreted crust, the crust and the
core.  The short vertical dotted lines in the left panel mark the
positions of the heat sources in the F+18 model. Panels \emph{c} and
\emph{d} show a zoom of selected curves from panels \emph{a} and
\emph{b}, respectively, near the crust/core interface.
}
\label{fig:rT100imp}
\end{figure*}
%%%%%%%%%%%%%%%%%%%%%%%%%%%%%%%%%%%%%%%%%%%%%%%%%%%%%%%%%

%%%%%%%%%%%%%%%%%%%%%%%%%%%%%%%%%%%%%%%%%%%%%%%%%%%%%%%%%
\begin{figure}
\centering
\includegraphics[width=\columnwidth]{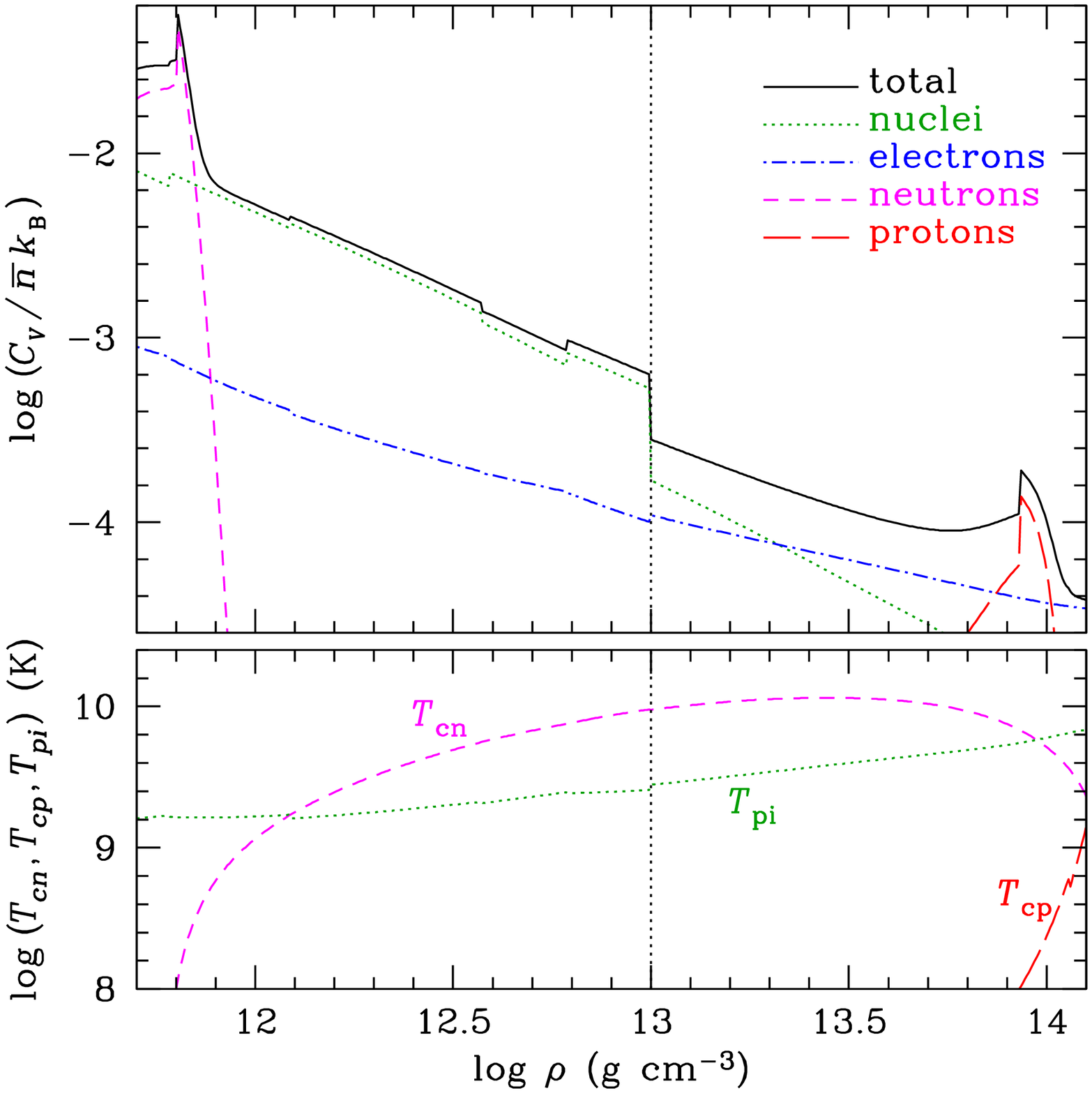}
\caption{\emph{Upper panel}: Heat capacity per one baryon in units of
$\kB$ as a function of density in the partially accreted inner crust at
$T=10^8$~K. Partial contributions from the nuclei (dotted line),
electrons (dot-dashed lines), dripped neutrons (short-dashed line), and
dripped protons (long-dashed line), as well as their sum (solid line)
are shown. \emph{Lower panel}: Critical temperatures of superfluidity of
neutrons (short-dashed lines) and protons (long-dashed lines), and the
ion plasma temperature (dotted line).
}
\label{fig:cvtc}
\end{figure}
%%%%%%%%%%%%%%%%%%%%%%%%%%%%%%%%%%%%%%%%%%%%%%%%%%%%%%%%%

Figure~\ref{fig:zmean} shows the charge number of a nucleon cluster
$Z_\mathrm{cl}$ at the center of a Wigner-Seitz sphere for four
energy density functionals. The filled symbols represent the calculated
values and the lines are the analytic fits for the ground state,
$Z_\mathrm{cl,gs}$, from \citet{Pearson_18}. The open symbols show the
mean charge number 
$
\langle Z_\mathrm{cl}\rangle\equiv \sum_Z Y_Z Z,
$
where $Y_Z$ is given by the approximation (\ref{E_exc}), (\ref{Y_Z})
with $e(\nbar,Z)$ taken from 
the electronic supplement to the paper by \citet{Pearson_18}.
 Despite the simplicity of approximation (\ref{Y_Z}),
 the evaluation of $Y_Z$ was not quite
 straightforward because of considerable numerical noise that had to
 be filtered-out from this supplement. Some density entries in the original
 tables have been omitted for this reason. 
 The mean values of nuclear mass and charge numbers and the impurity
parameter (\ref{Qimp}), estimated in the present work, will be
available in electronic
form at the CDS (\url{http://cdsarc.u-strasbg.fr/viz-bin/cat/J/A+A/}).
 
Models BSk24 and BSk26 predict the constant ground state charge number 
$Z_\mathrm{cl,gs}=40$ for almost entire inner crust.  Near the crust
bottom, $Z_\mathrm{cl,gs}$ starts to vary, because the stabilizing
effect of the shell corrections vanishes due to the proton drip.  In the
same deep layers, energy differences between different nuclei strongly
decrease (see Fig.~\ref{fig:energy24}), whence the deviations of
$\langle Z_\mathrm{cl}\rangle$ from $Z_\mathrm{cl,gs}$ arise. As shown
by \citet{Pearson_20}, at these high densities the true ground state may
become the pasta phase, instead of the phase of quasi-spherical nuclei.
However, the true ground state depends on small and not yet fully
determined pairing and shell corrections to the energies of rodlike and
slablike nuclei. We do not consider these exotic shapes hereafter. 

In contrast with the models BSk24 and BSk26, models BSk22 and BSk25
predict additional discontinuities of the charge number at lower
densities $\nbar<\npd$. The jumps of $Z_\mathrm{cl,gs}$ arise because
different local energy minima play the role of the absolute minimum at
different densities, as illustrated in Fig.~\ref{fig:energy25}. Around
the density where such a jump occurs, two local minima provide nearly
equal energy values, resulting in a mixture at a finite $T=\Ta$.
Accordingly, the equilibrium mean charge number $\langle
Z_\mathrm{cl}\rangle$ varies smoothly, instead of showing the
discontinuity, as we see in Fig.~\ref{fig:zmean}.  The corresponding
impurity parameter  $Q_\text{imp}$ [\req{Qimp}] is shown in
Fig.~\ref{fig:qimp}.  We see that the inner crust is rather pure in the
models BSk24 and BSk26 at densities $\nbar$ below the proton drip
density $\npd$.  Only at $\nbar>\npd$ the impurity parameter
$Q_\text{imp}$ rises by two orders of magnitude, because the stabilizing
effect of shell corrections vanishes. For the models BSk22 and BSk 25,
in contrast, there are layers with high $Q_\text{imp}$ also at 
$\nbar<\npd$, around the densities where two local minima of $E(Z)$ have
nearly equal depths, which leads to a mixture of  two types of nuclei
with different $Z$.

When the present work had been completed, a research by
\citet{Carreau_FG20} was published, where the impurity parameter in the
inner crust of a neutron star was calculated using a compressible liquid
drop description of the nuclei. The authors assumed that the nuclear
reactions stop at the point of Coulomb crystallization. We do not rely
on such assumption, because the above-mentioned ``freezing'' of nuclear
reactions does not necessarily coincide with the freezing of a Coulomb
liquid: at high temperatures ($T\gg10^9$~K), although heavy nuclei can
be arranged in a lattice due to the high pressure, they still may
undergo transformations (for example, through exchange of alpha
particles and free nucleons, which are not crystallized).  Therefore
$\Ta$ may differ from the Coulomb melting temperature.
\citet{Carreau_FG20} assumed that nuclear shell effects nearly vanish at
the temperature of crust formation, therefore they did not obtain the
peaked behavior of $Q_\mathrm{imp}$, which is seen in our
Fig.~\ref{fig:qimp}. Nevertheless, the results of \citet{Carreau_FG20}
are similar to our results by order of magnitude, showing an increase of
the impurity parameter from relatively low values at low densities to 
$Q_\mathrm{imp}$ about several tens near the bottom of the crust.

%%%%%%%%%%%%%%%%%%%%%%%%%%%%%%%%%%
\subsection{Conductivities in the crust with nuclear mixtures}
\label{sect:impconduct}

Figure \ref{fig:sigma} shows the electrical conductivities $\sigma$ as
functions of mass density $\rho$ in the pure nonaccreted crust and in
the crust with $Q_\text{imp}$ shown in Fig.~\ref{fig:qimp}. In agreement
with the conjecture by \citet{PonsVR13}, a layer with a low conductivity
appears near the crust bottom, if we allow for mixing the nuclei
according to their frozen equilibrium distribution. Note that an
assumption of non-spherical nuclei is not involved here. Moreover, there
are additional layers with strongly suppressed conductivity at lower
densities for the EoS models BSk22 and BSk25. The conductivity
depletions become more salient with the decrease of temperature.

Figure \ref{fig:kappa_imp} shows the thermal conductivities $\kappa$ as
functions of mass density $\rho$ at temperatures $10^7$~K and $10^9$~K,
calculated using the $Q_\text{imp}(\rho)$ dependences shown in
Fig.~\ref{fig:qimp}, compared with the conductivities computed, as in
Fig.~\ref{fig:kapt8}, with $Q_\text{imp}=0$. As well as in the case of
the electrical conductivities, $\kappa$ is suppressed near
the crust bottom for all four considered energy-density functionals, and
there are additional layers of low $\kappa$ at smaller $\rho$ in the
models BSk22 and BS25.

%%%%%%%%%%%%%%%%%%%%%%%%%%%%%%%%%%
\subsection{Heating and cooling of a crust with deep impurities}
\label{sect:coolimp}

The lowering of conductivities in the nonaccreted inner crust of a
neutron star due to the impurity distributions, evaluated above,  may
have observable effects on the short-term thermal evolution of the crust
during and after an outburst, if the accreted matter does not
fill the entire crust. In this case, the nonaccreted part of the crust
should contain the impure layers with reduced conductivities, left after
the initial neutron star cooling. To test this possibility, we performed
a series of short-term heating and cooling simulations. Some of the
computed lightcurves are shown in Fig.~\ref{fig:tT100imp}. This figure
illustrates the time dependence of the heat flux from the interior to
the surface, converted into $\kB\tTeff$, during and after the outburst
with the same parameters as in
Figs.~\ref{fig:tcky}\,--\,\ref{fig:r8vs18} for a neutron star with mass
$M=1.4\,M_\odot$, assuming that the accreted matter fills the crust to
the density $\rho_\mathrm{acc}=10^{13}$ \gcc. As is seen in
Fig.~\ref{fig:heatsrc}, this accreted crust is sufficiently thick to
include all layers with the most powerful heat release during an
outburst. Meanwhile, the
most impure layers with lowered conductivities 
at $\rho>10^{13}$ \gcc{} rest intact. To ensure the accreted crust of
such thickness and simultaneously provide the redshifted temperature of
the stellar core $\tilde{T}=10^8$~K, the long-term accretion should last
8.7 Myr at an average accretion rate
$\langle\dot{M}\rangle\approx10^{-10}\,M_\odot$ yr$^{-1}$. Figure
\ref{fig:tT100imp}  shows the results of simulations for two EoS models
BSk24 and BSk25. The lightcurves for the traditional model of pure
crust are compared with the models with impurity parameter shown in
Fig.~\ref{fig:qimp}. We see that the presence of nuclear mixtures in the
inner part of the crust delays relaxation. This effect has been
anticipated, because the relatively low thermal conductivity hampers
diffusion of the heat stored in the inner part of the crust. Naturally,
this effect is larger in the case of the BSk25 functional, because of an
additional layer with nuclear mixtures at moderate densities (see
Fig.~\ref{fig:qimp}). 

However, the relaxation delay due to the impurities is rather small,
despite the suppression of conductivities in
Fig.~\ref{fig:kappa_imp}. The reason for this smallness becomes clear by
looking at the evolution of temperature distribution in the crust, shown
in Fig.~\ref{fig:rT100imp}. In this figure, the two right panels present
a  zoom of selected curves from the two left panels to the inner part of
the crust, where the impurities are concentrated. We see the different
temperature distributions in the models with and without the impurities
in the deep layer of the crust, as expected for the strongly different
conductivities. However, temperature in these layers is anyway
substantially lower than the maximum, which corresponds to relatively
small heat stored in these layers during the outburst.

With or without impurities, most of the released heat leaks through the
deep crustal layers to the core, which plays the role of a thermostat
due to its large mass and high thermal conductivity. The smallness of
the heat that is stored near the crust bottom is explained by low heat
capacity, which is illustrated by Fig.~\ref{fig:cvtc}. Different
contributions to the heat capacity are evaluated as in \citet{PPP15}.
The heat capacity per baryon $C_V/\nbar$ decreases from the top to the
bottom of the inner crust by a factor of several tens. The contribution
of the crystalline lattice of the nuclei (ions) 
$C_{V,\mathrm{i}}/\nbar\kB$ decreases with the increase of the ion
plasma temperature
$T_\mathrm{pi}\propto[\rho\,Z_\mathrm{cl}^2/(A'A)]^{1/2}$ and the number
of baryons per Wigner-Seitz cell $A'$ as $C_{V,\mathrm{i}}/\nbar \propto
(T/T_\mathrm{pi})^3/A'$ in the strong quantum limit $T\ll
T_\mathrm{pi}$. The electron heat capacity
$C_{V,\mathrm{e}}\propto\nbar(Z^2/A')\,T/T_\mathrm{F}$ is
suppressed due to the strong degeneracy
($T\ll T_\mathrm{F}$, where
$T_\mathrm{F}=(\epsilon_\mathrm{F}-\mel c^2)/\kB$ is the electron Fermi
temperature).
The contributions of dripped nucleons are suppressed because of their
superfluidity (see, e.g., \citealt{YakovlevLS99}). The critical
temperatures of neutron and proton superfluidity ($T_\mathrm{cn}$ and
$T_\mathrm{cp}$) and the ion plasma temperature ($T_\mathrm{pi}$) are
shown in the lower panel of Fig.~\ref{fig:cvtc}. Nevertheless, despite
the smallness of the additional stored heat, the presence of the
impurities can help to agree theoretical models with observations, as we
will see in the next section.

%%%%%%%%%%%%%%%%%%%%%%%%%%%%%%%%%%
\section{MXB 1659--29}
\label{sect:MXB}

In this section we apply the theoretical models, described above, to the
SXT MXB 1659$-$29 (MAXI J1702$-$301). This eclipsing quasi-persistent
transient was discovered as an X-ray bursting source during the \textit{SAS-3}
satellite mission \citep{LewinHD76,LewinJoss77}. It has been observed
many times using different instruments (see
\citealt{Wijnands_03,Parikh_19}, and references therein). The source was
detected several times in X-rays  and in the optical from October 1,
1976 till July 2, 1979, but then (before July 17, 1979;
\citealt{Cominsky_83}) it turned into quiescence and could not be
detected any more until April 1999, when \citet{intZand_99} reported it
to be active again. The source remained bright for almost 2.5 yr before
it became dormant again in September 2001. It was first observed in the
quiescent state using \textit{Chandra} in October 2001
\citep{Wijnands_03}. Afterwards the thermal emission powered by the
crust cooling of the neutron star in this SXT was observed several times
till October 2012 \citep{Cackett_06,Cackett_08,Cackett_13}. In August
2015 the source showed a new outburst \citep{Negoro_15}, which lasted
$\approx550$ days till February 2017. Subsequent crust cooling was
followed from March 2017 using X-ray observatories \textit{Swift},
\textit{Chandra}, and \textit{XMM-Newton}. The results have been
summarized and analyzed by \citet{Parikh_19}. Following these authors,
we name \emph{outburst I} and \emph{outburst II} those of 1999\,--\,2001
and 2015\,--\,2017, respectively. In addition we name \emph{outburst 0}
the one observed in 1976\,--\,1979. \citet{Wijnands_03} point out that
the source might have also been detected during the period 1971 to 1973
using \textit{Uhuru} (classified as 4U 1704$-$30; \citealt{Forman_78}),
but this identification with MXB 1659$-$29 is not certain.

Analyzing observations of MXB 1659$-$29 performed in July 2012 (the last
ones before outburst II) \citet{Cackett_13} discovered that the count
rate has dropped in the latest observation compared with the previous
two, taken approximately 4 and 7 years earlier, although the effective
temperature remained at the same level (within uncertainties). Inclusion
of a power-law component, in addition to the thermal component of the
spectrum, improved the fit and gave a significantly (by $\sim20$\%)
lower effective temperature, but a reanalysis of the previous
observations showed that they do not require the power-law component.
Another possible explanation suggested by \citet{Cackett_13} was an
increase in the column density on the line of sight between these
observation.

%%%%%%%%%%%%%%%%%%%%%%%%%%%%%%%%%%
\subsection{Modeling the observed crust cooling}

The quiescent lightcurves after the end of outburst I have been modeled
previously in a number of works
\citep{BrownCumming09,Cackett_13,Deibel_17,Parikh_19}. The theoretical
models can only fit the observations if one includes so called ``shallow
heating'' at $\rho\sim10^8-10^{10}$ \gcc{} in addition to the deep
crustal heating predicted by the \citet{HZ90,HZ08} theory, with extra
heat deposited at the outer crust densities.  The shallow heating is
necessary for consistency of the theory and observations not only in MXB
1659$-$29, but also in the other SXTs that show the crust cooling (see
the review by \citealt{WijnandsDP17} and Table~I in \citealt{Chamel_20}). 
The origin of this shallow heat
is unclear. For example, it may be related to the heat deposited in the
outer crust by the nuclear reactions that power the X-ray bursts
observed during the active phase of accretion (see \citealt{Meisel_18},
for review). 

\citet{Parikh_19} presented observations and consistent modeling of the
short-term evolution of MXB 1659$-$29 during and after the two 
outbursts I and II. In the analysis of the observations and in the
numerical simulations the authors assumed neutron star mass
$M=1.6\,M_\odot$ and radius $R=12$ km. The authors used the neutron star
heating and cooling code \texttt{NSCool} \citep{NSCool}, which employs
simplifying assumptions of a quasi-stationary blanketing envelope at
densities $\rho<\rhob$ (with $\rhob=10^8$ \gcc{} in their case) and the
barotropic EoS at higher densities (we examined  the inferences of such
assumptions on the computed $\tTeff(t)$ in Sect.~\ref{sect:CKY} above).
The amount of light elements in the envelope was allowed to freely vary.
The impurity parameters $Q_\text{imp}$ were allowed to vary
independently in the outer crust, an upper part of the inner crust, and
a bottom layer of the inner crust. Moreover, either the envelope
composition or the shallow heating parameters, viz.{} the energy
$E_\mathrm{sh}$ deposited per an accreted baryon in a shallow layer and
the depth of this layer, were allowed to vary between the outbursts I
and II. The best fits were obtained for iron blanket replaced by light
elements to the mass density $\sim 10^{5}-10^{6}$ \gcc, shallow heat
$E_\mathrm{sh}\sim1.2\pm0.7$ MeV at $\rho$ between $10^8$ and $10^{10}$
\gcc, and $Q_\text{imp}\sim2$.

We model the short-term thermal evolution of a neutron star during and
between the outbursts consistently with its long-term evolution.  The
latter is modeled in the same way as in Paper~I. 
As well as in Sects.~\ref{sect:CKY}
and~\ref{sect:coolimp}, we compute thermal structure and evolution of
the star without the assumption that the EoS is barotropic. For this
purpose we employ the code described in \citet{PC18}. This allows us to
get rid of a thick quasi-stationary heat-blanketing envelope, required
in the standard neutron-star cooling codes (such as CKY
or \texttt{NSCool}), and to treat
evolution of non-degenerate and partially degenerate envelopes on equal
footing with the strongly degenerate interior of the star. In
Sect.~\ref{sect:CKY} we have seen that it can be important for
reproducing the early stage ($t\lesssim0.1$ yr) of crust cooling. To
test the potential effect of the highly impure layers near the bottom of
the crust, we assume that the accreted matter has replaced the
ground-state matter only partially, down to $\rho_\mathrm{acc}=10^{13}$
\gcc. Instead of the HZ'08 model of the deep crustal heating, we use the
more recent F+18 model ($E_\mathrm{h}=1.44$ MeV per baryon for
$\rho_\mathrm{acc}=10^{13}$ \gcc, which is already close to
$E_\mathrm{h}=1.54$ MeV per baryon for the fully accreted crust).

To compare theoretical heating and cooling
curves with observations, we mostly rely on the observational data
presented by \citet{Parikh_19}. For the observation of 2012, which was
discarded by these authors because of the above-mentioned uncertainties
in its spectral fitting, we plot both estimates
$k\tilde{T}_\mathrm{eff}=43\pm5$ eV and $55\pm3$ eV obtained by 
\citet{Cackett_13} with and without inclusion of the power-law spectral
component.

For a given neutron-star model, the quasi-equilibrium redshifted 
luminosity in quiescence $\tilde{L}_\mathrm{eq}$ or, equivalently, the
quasi-equilibrium effective temperature $\tilde{T}_\mathrm{eff,eq}$, is
determined by the accretion history. Effectively
$\tilde{T}_\mathrm{eff,eq}$ is mainly determined by the mean accretion
rate $\langle\dot{M}\rangle$ during the preceding $\sim10$ kyr. Since
the available observations cover much shorter intervals $\Delta
t\lesssim50$ yr, the estimates of $\langle\dot{M}\rangle$ are rather
uncertain. The mean accretion rate is usually evaluated as
$\langle\dot{M}\rangle\sim M_{\Delta t}/\Delta t$, where $M_{\Delta
t}=\int_{t-\Delta t}^t \dot{M}\dd t$ is the total mass accreted during
the period of observations $\Delta t$, and $\dot{M}(t)$ is the
instantaneous outburst accretion rate, which is related to the
bolometric outburst  luminosity $\tilde{L}(t)$ by equation
\citep[e.g.,][]{Meisel_18}
\bea
   \dot{M}(t) &=& \tilde{L}(t) \frac{(1+\zg)^2}{c^2\zg}
     = \left(1+\frac{\zg}{2}\right)\frac{\tilde{L}(t) R}{G M}
\nonumber\\
     &\approx& 1.2\times10^{-9}\left(1+\frac{\zg}{2}\right)
     \frac{L_{37}R_{10}}{M/M_\odot}
     \,M_\odot \,\mbox{yr}^{-1},
\label{LXvsMdot}
\eea
where $L_{37}\equiv\tilde{L}(t)/10^{37}$ erg s$^{-1}$ and
$R_{10}\equiv R/10$ km.
For a neutron star with $M\sim1-2\,M_\odot$ and $R\sim10$ km, one has
$\dot{M}\sim5 \tilde{L}/c^2$ \citep[e.g.,][]{VanIH19}.

In the case of MXB 1659$-$29, the bolometric flux during outbursts I is
on the average $\sim3\times10^{-9}$ erg cm$^{-2}$ s$^{-1}$
\citep{Parikh_19}. By an analysis of several X-ray bursts during
outburst I, \citet{Galloway_08} derived the distance estimates $9\pm2$
or $12\pm3$ kpc, depending on the assumed thermonuclear fuel
composition. For a neutron star with $M\sim1.6\,M_\odot$ and $R\sim12$
km this leads, according to \req{LXvsMdot}, to the accretion rates
$\dot{M}\sim(4\pm2)\times10^{-9}\,M_\odot$ yr$^{-1}$ during outburst I.
The outburst II shows a strong (up to an order of magnitude)
variability, being approximately three times weaker than outburst~I on
the average \citep{Parikh_19}. An average accretion rate of
$4\times10^{-9}\,M_\odot$ yr$^{-1}$ during the 2.5 years of outburst I
gives the accreted mass of $10^{-8}\,M_\odot$, and outburst II adds
$\sim20$\% to this value. Taking the base $\Delta t=37.6$ yr from the
end of outburst 0 to the end of outburst II, we obtain
$\langle\dot{M}\rangle\sim3\times10^{-10}\,M_\odot$ yr$^{-1}$, which is
somewhat larger than the earlier estimate
$\langle\dot{M}\rangle=1.7\times10^{-10}\,M_\odot$ yr$^{-1}$
\citep{Heinke_07}, used as a fiducial value up to now
(e.g., \citealt{WijnandsDP17} and Paper~I). Alternatively,
assuming that outburst~0 had the same intensity and duration as
outburst~I and using the timespan $\Delta t=38.9$ yr from the first
detection of outburst 0 to the start of outburst II, we obtain
$\langle\dot{M}\rangle\sim10^{-9}\,M_\odot$ yr$^{-1}$. Using the total
timeline of observations from 1976 to 2020, we arrive at 
$\langle\dot{M}\rangle\sim5\times10^{-10}\,M_\odot$ yr$^{-1}$. Taking
all the uncertainties into account, we consider
$\langle\dot{M}\rangle\sim(10^{-10}-10^{-9})\,M_\odot$ yr$^{-1}$ as
compatible with observations.

The long-term thermal evolution of a neutron star depends on the baryon
superfluidity in its core. In this section, for the neutron triplet-type
pairing gap in the core we use the parametrization of \citet{Ding_16} to
their  computations based on the N3LO Idaho potential with many-body
correlations taken into account. For the proton singlet-type
superfluidity we use the BS  parametrization of \citet{Ho_15} based on a
theoretical model computed by \citet{BaldoSchulze07}. 
Following
\citet{Ho_15} (also see \citealt{LevenfishYakovlev94}), we applied
different coefficients of conversion from the zero-temperature gap
energy to the critical temperature for the singlet and triplet
superfluidity types. Then the maximum critical temperatures in the core
are  $T_\mathrm{cn}=2.4\times10^8$~K at $\rho=5\times10^{14}$ \gcc{}
and  $T_\mathrm{cp}=4.7\times10^9$~K at $\rho=2\times10^{14}$ \gcc.

%%%%%%%%%%%%%%%%%%%%%%%%%%%%%%%%%%%%%%%%%%%%%%%%%%%%%%%%%
\begin{figure*}
\centering
\includegraphics[width=\columnwidth]{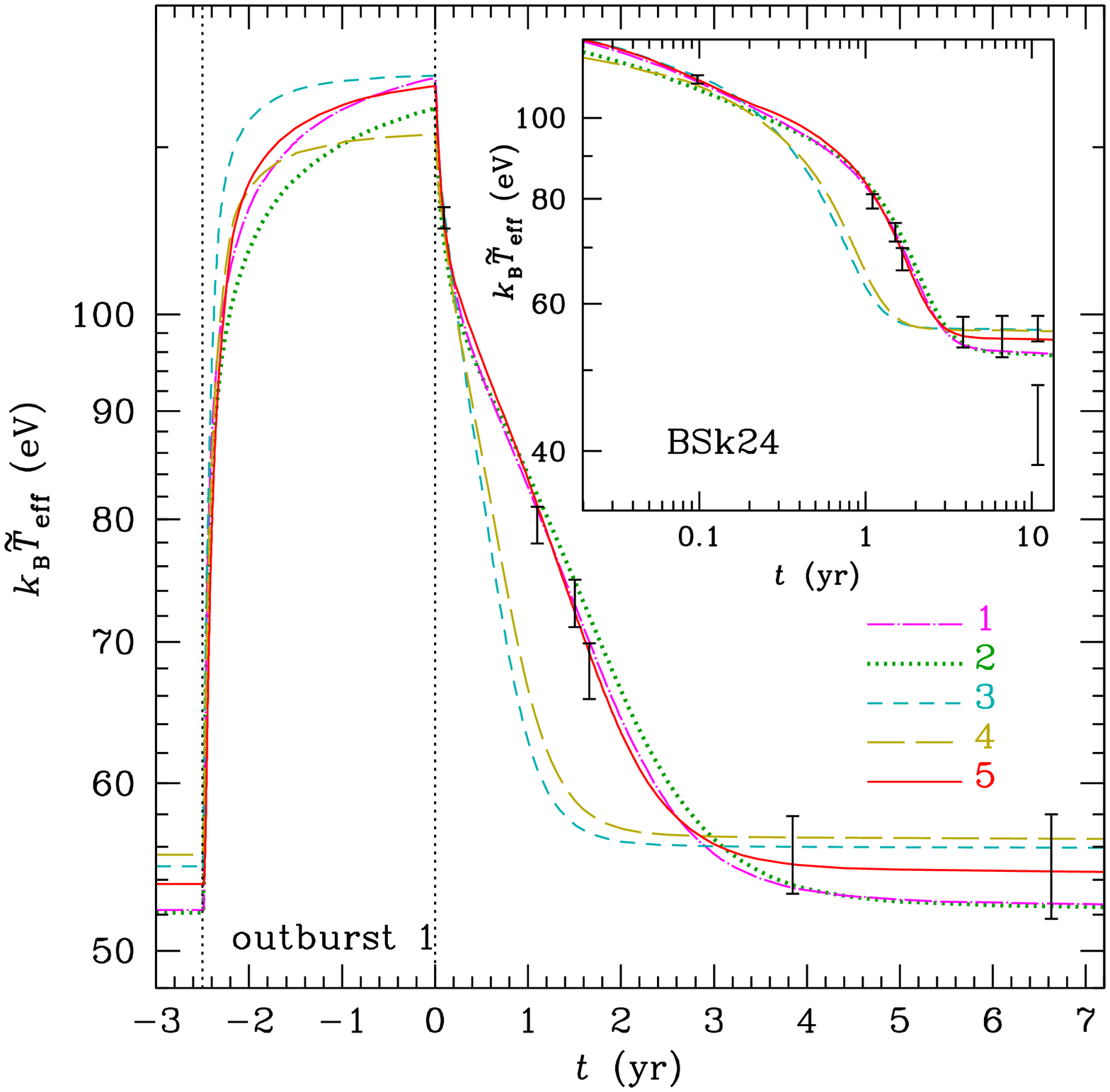}
\includegraphics[width=.561\columnwidth]{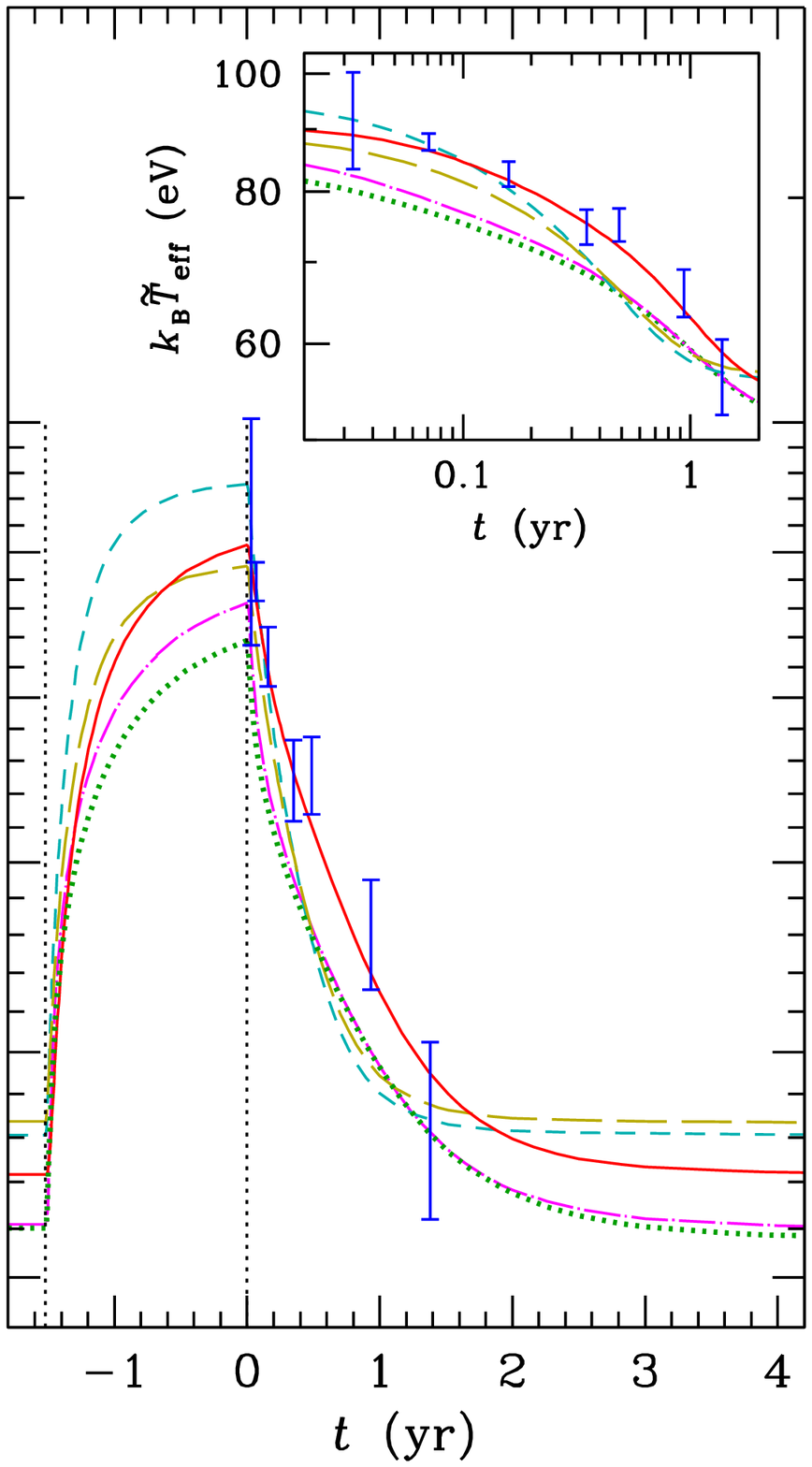}
\caption{Simulated lightcurves for the outbursts I (left panel) and II 
(right panel) of MXB 1659$-$29 versus observations. The lightcurves
have been computed using the BSk24 EoS model for a neutron star with
$M=1.65\,M_\odot$ with a partially accreted envelope under different
assumptions on the envelope and crust composition. The envelope is
composed either of carbon (curves 1, 2) or of helium (curves 3, 4, 5).
The accreted crust is modeled according to F+18 and is assumed to extend
down to $\rho_\mathrm{acc}=10^{13}$ \gcc. Beyond this density, the
nonaccreted crust is either in ground-state (curves 1, 3) or in frozen
equilibrium state (curves 2, 4, 5). The accreted crust is either pure
(curves 1\,--\,4) or not (curve 5). The errorbars show the spectral
fitting results from \citet{Parikh_19} (at the 90\% confidence), except
for the two bars at $t=10.8$ yr in the left-panel inset, which represent
two alternative fits from \citet{Cackett_13}. In the numerical
simulations, the long-term mean accretion rate is adjusted to the
equilibrium value within uncertainties, and the shallow heating is
adjusted so as to reproduce the first observation in quiescence after
outburst~I, within uncertainties. The parameters of the simulations are
listed in Table~\ref{tab:models} (see the text for details).
}
\label{fig:MXB24}
\end{figure*}
%%%%%%%%%%%%%%%%%%%%%%%%%%%%%%%%%%%%%%%%%%%%%%%%%%%%%%%%%

\begin{figure}
\centering
\includegraphics[width=\columnwidth]{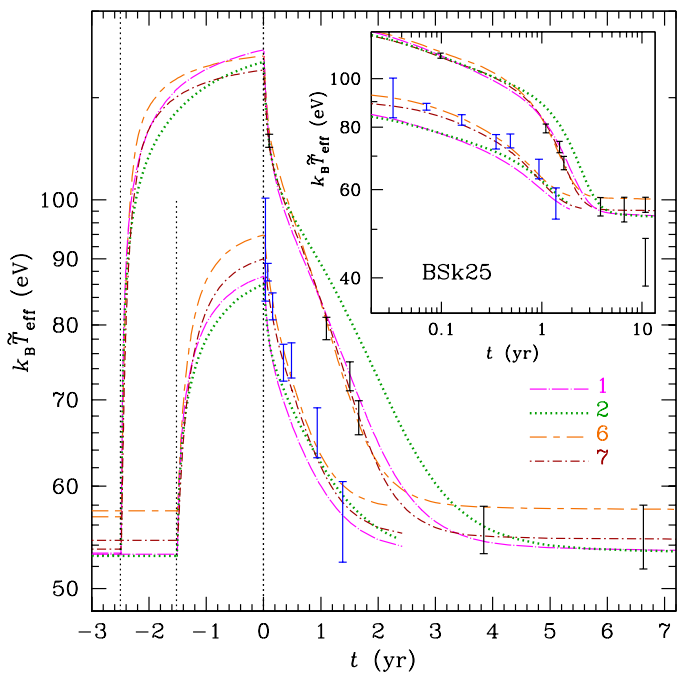}
\caption{Simulated lightcurves for the outbursts I and II (upper and
lower lines and errorbars, respectively) versus observations, as in
Fig.~\ref{fig:MXB24} but using the BSk25 EoS model. Curves 1 and 2 are
computed assuming carbon envelope and the same accreted crust
composition and heating parameters as in Fig.~\ref{fig:MXB24}; curves 6
and 7 are computed assuming helium envelope with the parameters adjusted
to observations as in Fig.~\ref{fig:MXB24} (see Table~\ref{tab:models}
and the text for
details).
}
\label{fig:MXB25}
\end{figure}

In the F+18 model of the accreted crust for a standard neutron star
cooling (mainly by the modified Urca processes), the above-mentioned
estimates of the mean accretion rate of MXB 1659$-$29 correspond to the
quasi-equilibrium redshifted effective temperature
$\kB\tTeff\sim(70-100)$ eV, much higher than the one inferred from
observations. A tentative explanation that only a small part of the
crust is accreted and affects the heating, which we considered in
Paper~I, is hard to agree with the observed relaxation time of the order
of several years: in this case the post-outburst relaxation had to be
much faster (see Sect.~\ref{sect:CKY}). A more plausible explanation of
the low observed temperature is the enhanced cooling due to the direct
Urca process, which is possible if the star mass exceeds certain
threshold $M_\mathrm{DU}$ \citep[e.g.,][]{Haensel95}. In
Figs.~\ref{fig:MXB24} and~\ref{fig:MXB25} we compare the observational
data with simulations performed using the BSk24 and BSk25 EoS models for
a neutron star with mass $M=1.65\,M_\odot$, which is slightly above the
direct Urca thresholds $M_\mathrm{DU}\approx1.6\,M_\odot$ for these
models. The gravitational redshifts in these models are close to
$\zg\approx0.28$, implied in the spectral analysis by \citet{Parikh_19}
who assumed $M=1.6\,M_\odot$ and $R=12$ km. 

The simulations were performed as follows. First we  simulate a
long-term evolution with the accretion rate equal to the assumed average
$\langle\dot{M}\rangle$ during the time required to fill the crust by
the replaced matter to $\rho_\mathrm{acc}=10^{13}$ \gcc. This
establishes the temperature of the core. Next, to bring the crust to
equilibrium, we continue the modeling during the next 20 years without
accretion. Then we model the heating during outburst 0 that ended in
1979. In the absence of detailed observational information on this early
outburst, we assume that it is similar to outburst~I. We fix
$\dot{M}=4\times10^{-9}\,M_\odot$ yr$^{-1}$ and duration 2.5 years for
these outbursts, and 20 years of quiescence between them. After the end
of outburst~I we trace the 13.9 years of cooling in quiescence.
Afterwards outburst~II proceeds during 1.52 yr at
$\dot{M}=1.3\times10^{-9}\,M_\odot$ yr$^{-1}$, and finally we trace the
cooling after the end of outburst~II.
We computed the short-term thermal evolution of
the entire star, including its outer envelope, crust, and core
simultaneously, without assumption of a quasi-stationary 
structure of the core or the envelope.

Let us first discuss the results shown in Fig.~\ref{fig:MXB24}.  Here,
the BSk24 EoS model is used, which implies a relatively thin layer with
a high impurity content near the crust bottom (see
Sect.~\ref{sect:impcrust}). For the mass $M=1.65\,M_\odot$, stellar
radius is $R=12.6$~km, and the total accretion time needed to fill the
crust to the assumed density $\rho_\mathrm{acc}=10^{13}$ \gcc{} is
$t_\mathrm{acc}\approx(7.6/\langle\dot{M}\rangle_{-10})$~Myr, where
$\langle\dot{M}\rangle_\mathrm{-10} \equiv
\langle\dot{M}\rangle/(10^{-10}\,M_\odot$ yr$^{-1}$). We adjust
$\langle\dot{M}\rangle$ so as to reach, within uncertainties, the
observed quasi-equilibrium thermal state of the neutron star in
quiescence. Our models could not reproduce the drop in luminosity
between observations of 2008 and 2012 implied by the spectral fit
including a power-law component \citep{Cackett_13}. Therefore we adopt
the result of modeling without additional power-law component, which
implies that the apparent fading is due to an increased absorption. It
is consistent with the previous two observations and gives
$\kB\tTeff=55\pm3$ eV. If the heat blanketing envelope of the neutron
star were made of iron, this equilibrium value would imply too high
temperature of the core, which could not be provided by realistic rates
of long-term accretion within our neutron-star model.  With carbon
envelope, the observational equilibrium level is reached near the upper
end $\langle\dot{M}\rangle=10^{-9}\,M_\odot$ yr$^{-1}$ of our allowed
range of accretion rates.  With helium envelope, it is reached near the
lower end of the range.

In agreement with the previous studies of MXB 1659$-$29, our simulations
show that a shallow heating should be included into the model. Otherwise
the rise of the effective temperature $\tTeff$ by the end of the
outburst would be much smaller than observed. The location of an
additional heating source may vary in the wide range of densities
$\rho\sim(10^8-10^{10})$ \gcc{} (e.g., \citealt{Parikh_19}). To be
specific, we add it to the most shallow source in the F+18 model at
$\rho\approx1.4\times10^9$ \gcc. The additional energy deposited per
each accreted baryon,  $E_\mathrm{sh}$, is the model parameter. In the
examples shown in Fig.~\ref{fig:MXB24} it is chosen to reproduce
the highest $\tTeff$ observed for this source in quiescence (the
\textit{Chandra} observation on October 15, 2001). We list the
parameters of the presented numerical models in Table~\ref{tab:models}.

Line 1 in Fig.~\ref{fig:MXB24} shows the best results obtained in
simulations without any impurities in the crust, performed  assuming the
carbon blanketing envelope. Although the simulation of all three
outbursts was performed in a single run for each parameter set, the time
on the horizontal axis is measured separately from the end of
outbursts~I and~II for convenience of presentation. The left panel of
Fig.~\ref{fig:MXB24} shows the results for outburst~I. By construction,
each simulated lightcurve agrees with the hottest observed point at
$t\approx1$ month after the first non-detection of outburst~I (this level
is regulated by $E_\mathrm{sh}$) and the near-equilibrium points at
$t\gtrsim4$ years (regulated mainly by $\langle\dot{M}\rangle$). The
intermediate observations show a satisfactory agreement with model~1.
The best-fit $E_\mathrm{sh}\sim700$ keV is within the range evaluated in
\citet{Parikh_19}.

%%%%%%%%%%%%%%%%%%%%%%%%%%%%%%%%%% 
\begin{table}
\centering
\caption[]{Parameters of modeling thermal evolution of MXB 1659$-$29:
sequential number in Figs.~\ref{fig:MXB24} and~\ref{fig:MXB25}, the
composition of the heat blanketing envelope, assumed impurity parameter
$Q_\text{imp}$ in the accreted crust, the model of the nonaccreted
crust bottom (pure ground state or equilibrium composition according to
Sect.~\ref{sect:impcrust}), mean long-term accretion rate $\langle
\dot{M}\rangle_\mathrm{-10} \equiv
\langle\dot{M}\rangle/(10^{-10}\,M_\odot$ yr$^{-1}$), and shallow
heat per accreted baryon $E_\mathrm{sh}$.
\label{tab:models} 
}
\begin{tabular}{cccccc}
\hline\hline
\noalign{\smallskip}
  &  envelope & $Q_\text{imp}$ & crust &
 $\langle\dot{M}\rangle_{-10}$ & $E_\mathrm{sh}$ \\
 no.& type & at $\rho<\rho_\mathrm{acc}$ & bottom &
 & (keV) \\
\hline                                                                                                                                                                            
\noalign{\smallskip}
  1 & C & 0 & pure & 10  & 700 \\
  2 & C & 0 & impure & 10 & 600 \\
  3 & He & 0 & pure & 1 &300 \\
  4 & He & 0 & impure & 1& 200 \\
  5 & He & 3 & impure & 2& 230 \\
  6 & He & 1 & impure & 2& 230 \\
  7 & He & 2 & impure & 1& 200 \\
\noalign{\smallskip}                                                                                                                                     
\hline\hline
\end{tabular}
\end{table}
%%%%%%%%%%%%%%%%%%%%%%%%%%%%%%%%%% 

A comparable but somewhat less good agreement is seen for model~2, which
differs from model~1 by the composition of the innermost (nonaccreted)
part of the crust, which is
now taken from the frozen equilibrium model of
Sect.~\ref{sect:impcrust}. The low-conductivity
layer causes a delay of crust cooling.
Since this layer is deep and thin, its
heat capacity is low (cf.{} Fig.~\ref{fig:cvtc}), hence it cannot
accumulate much heat. Therefore the cooling delay is rather
small also. On the other hand, this deep impure layer
slows down the leakage of the heat to the core during the outburst.
As a result, the same rise of temperature can be provided by a smaller
value of $E_\mathrm{sh}$.

A replacement of the carbon envelope by a more heat-transparent helium
envelope (lines  3\,--\,5 in Fig.~\ref{fig:MXB24}) has two immediate
effects. First, a lower core temperature is sufficient to give the
observed equilibrium level of the effective temperature. A lower
long-term accretion rate
$\langle\dot{M}\rangle\sim(1-2)\times10^{-10}\,M_\odot$ yr$^{-1}$, which
is close to the traditionally accepted one, suffices for that (the fifth
column of Table~\ref{tab:models}). Second, the heat that is stored in
the crust during the outburst is radiated away much quicker through the
transparent envelope. For this reason, the pure crust model (curve 3 in
Fig.~\ref{fig:MXB24}) becomes incompatible with observations. The bottom
layer with large  $Q_\text{imp}$ causes only a small delay in the crust
cooling (line~4), insufficient to agree the model with observations. To
reach the agreement, we have to assume that the accreted crust also
contains some impurities. The value $Q_\text{imp}=3$ provides the best
fit (line~5). The needed ``shallow heat'' is strongly reduced: the best
fit is obtained with $E_\mathrm{sh}=230$ keV.

The right panel of Fig.~\ref{fig:MXB24} shows a comparison of
theoretical lightcurves with observations for the crust cooling after
outburst~II. All the model parameters are kept the same as for
outburst~I, without any additional adjustment. We see that model~5
provides the best agreement with the data for this outburst also.

%%%%%%%%%%%%%%%%%%%%%%%%%%%%%%%%%%%%%%%%%%%%%%%%%%%%%%%%%
%%%%%%%%%%%%%%%%%%%%%%%%%%%%%%%%%%%%%%%%%%%%%%%%%%%%%%%%%

%%%%%%%%%%%%%%%%%%%%%%%%%%%%%%%%%%%%%%%%%%%%%%%%%%%%%%%%%
\begin{figure*}
\centering
\includegraphics[width=\columnwidth]{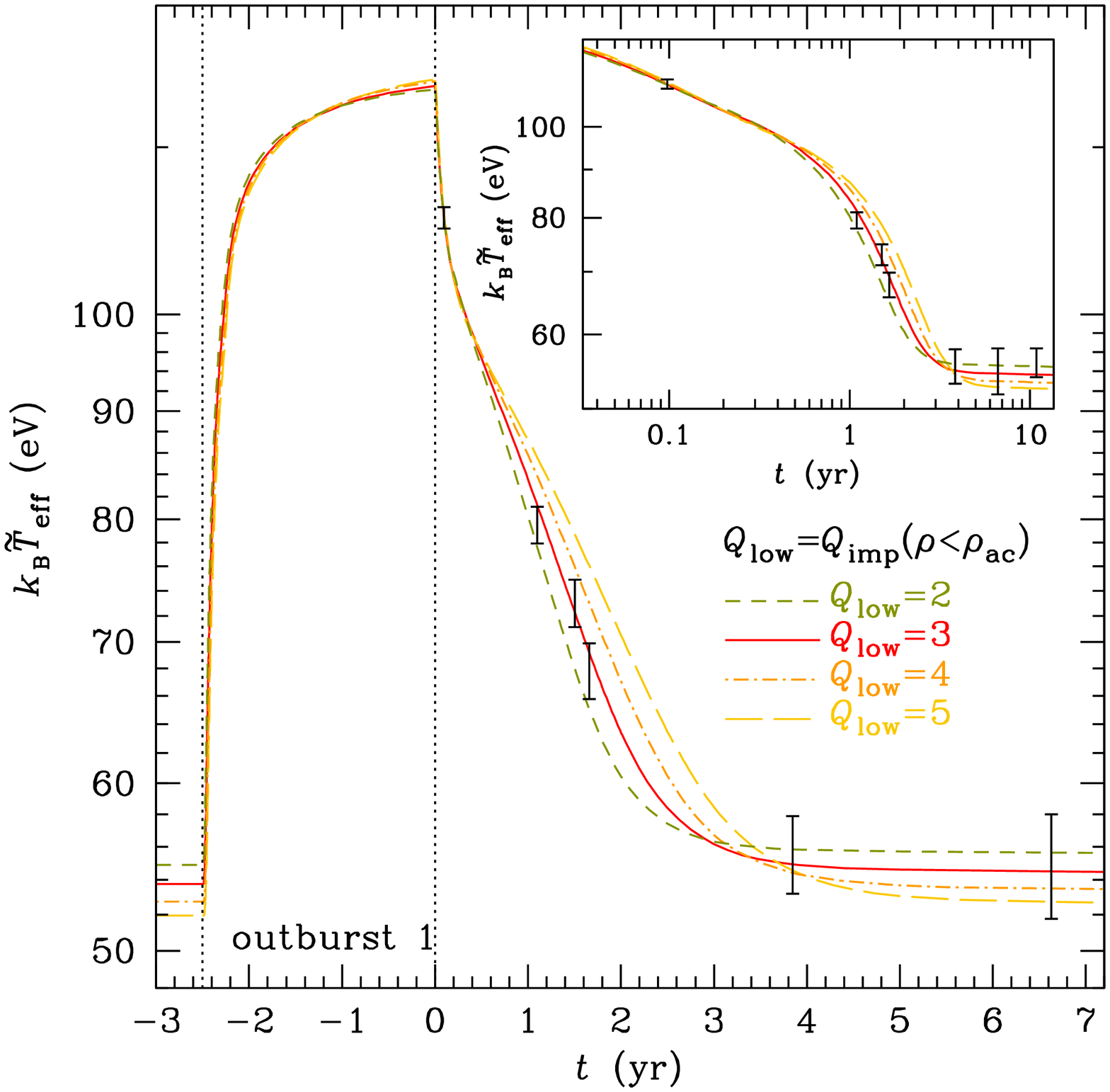}
\includegraphics[width=.892\columnwidth]{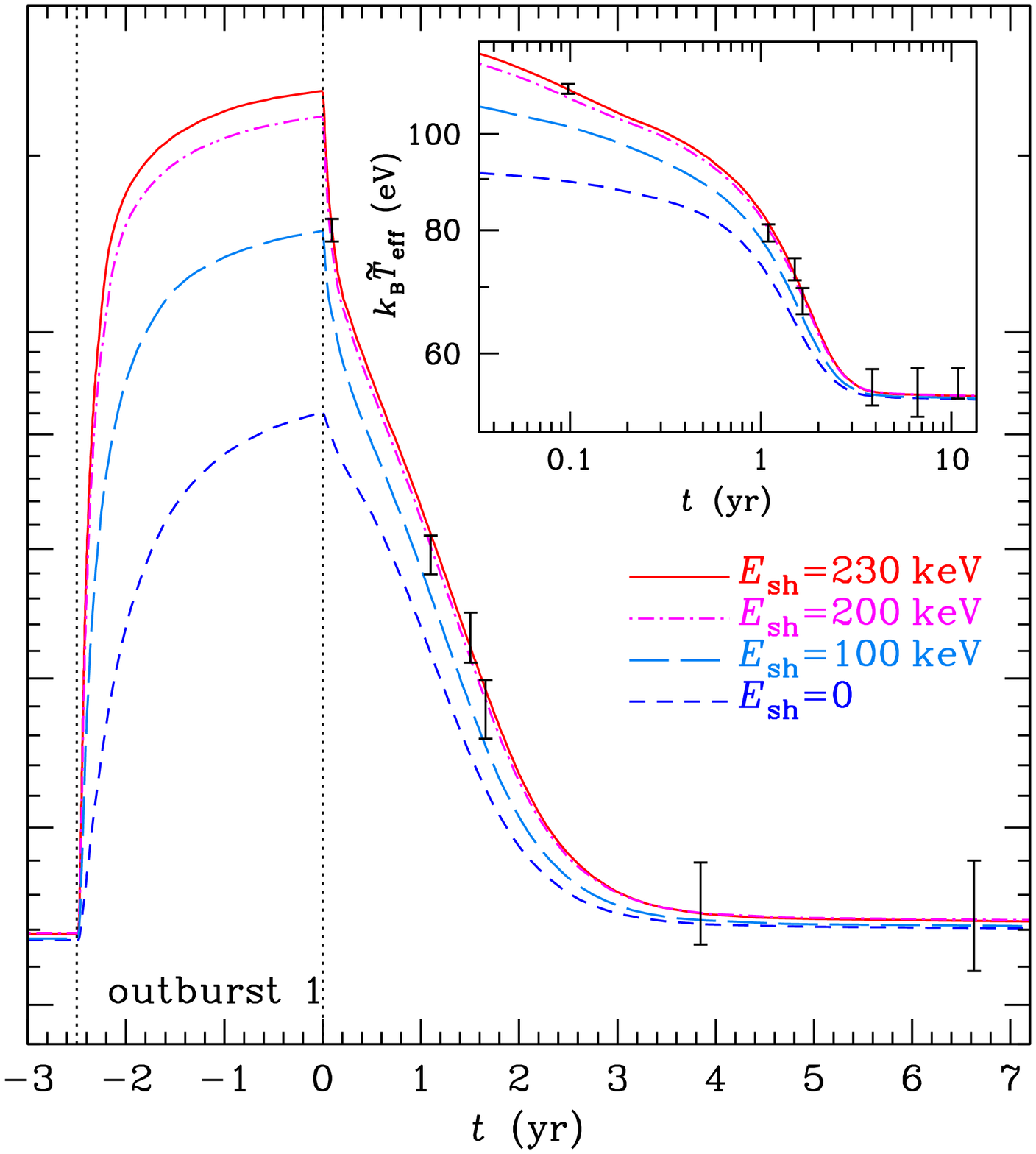}
\caption{Variations of simulated lightcurves
with varying impurity parameter $Q_\mathrm{imp}$ in the
accreted crust (left panel) or varying shallow heat $E_\mathrm{sh}$
(right panel). In both panels, errorbars reproduce the observations at
the cooling stage after outburst~I and the solid red curve is
the best-fitting curve 5 from the left panel of Fig.~\ref{fig:MXB24}.
Other curves show the lightcurves with increased or decreased 
$Q_\mathrm{imp}$ or $E_\mathrm{sh}$, according to the legend.
}
\label{fig:variations}
\end{figure*}
%%%%%%%%%%%%%%%%%%%%%%%%%%%%%%%%%%%%%%%%%%%%%%%%%%%%%%%%%

Figure~\ref{fig:MXB25} presents a comparison of the observational crust
cooling data with theoretical models for the EoS BSk25. In this case,
$R=12.4$~km, and the total accretion time needed to fill the crust
to $\rho_\mathrm{acc}=10^{13}$ \gcc{} is
$t_\mathrm{acc}\approx(7.9/\langle\dot{M}\rangle_{-10})$~Myr. We have
seen in Sect.~\ref{sect:impconduct} that additional depressions of
thermal conductivities appear in the BSk25 model, besides the depression
near the crust bottom. To see a direct influence of these additional
depressions, curves 1 and 2 in Fig.~\ref{fig:MXB25} are calculated with
the same model parameters (Table~\ref{tab:models}) as in
Fig.~\ref{fig:MXB24}. We see that model 1 provides a similar agreement
with the data for outburst~I, but noticeably underestimates $\tTeff$ at
$t\lesssim1$ yr after outburst~II. In model 2, on the contrary, the
decline of $\tTeff$ is significantly delayed (as anticipated because of
the slowdown of the heat diffusion in the inner crust) and becomes
longer than the observed one.

We do not show the models for pure accreted crust with helium envelope,
which are similar to models 3 and 4 in Fig.~\ref{fig:MXB24} and disagree
with observations. A satisfactory agreement is reached while assuming
$Q_\text{imp}\sim1$\,--\,2 (lines 6 and 7) in the accreted crust.
Both outbursts I and II are satisfactorily fitted using the same model
parameter sets. The best fit (line 7) as well as in the case of BSk24
(line 5 in Fig.~\ref{fig:MXB24}), requires a smaller ``shallow heat''
($E_\mathrm{sh}\approx200$ keV) compared with the case of a less
transparent carbon envelope ($E_\mathrm{sh}\approx600$ keV).
The deep layer with mixtures allows us to reduce $E_\mathrm{sh}$ by
$\sim100$ keV in both cases of the carbon and helium envelopes.

%%%%%%%%%%%%%%%%%%%%%%%%%%%%%%%%%%
\subsection{Relation of model parameters to lightcurve properties}

The results of the presented simulations of thermal evolution allow us
to trace the influence of the input parameters of the model on the
lightcurve of the crust relaxation for each assumed EoS of the
nonaccreted and accreted neutron-star matter, baryon superfluidity
models, and model of distribution of the deep crustal heat sources.\
First, the
mass of the star, thickness and composition of the accreted envelope,
and mean long-term accretion rate determine the limiting inter-outburst
equilibrium luminosity, as described in Paper~I. Second, with the
previous parameters being fixed, the impurity content of the accreted
crust mainly determines the average slope of the crustal cooling curve.
Third, the shallow
heat mainly determines the height of the peak of the thermal flux from
the crust to the surface, which corresponds to early post-outburst
relaxation. 

%%%%%%%%%%%%%%%%%%%%%%%%%%%%%%%%%%%%%%%%%%%%%%%%%%%%%%%%%
\begin{figure}
\centering
\includegraphics[width=\columnwidth]{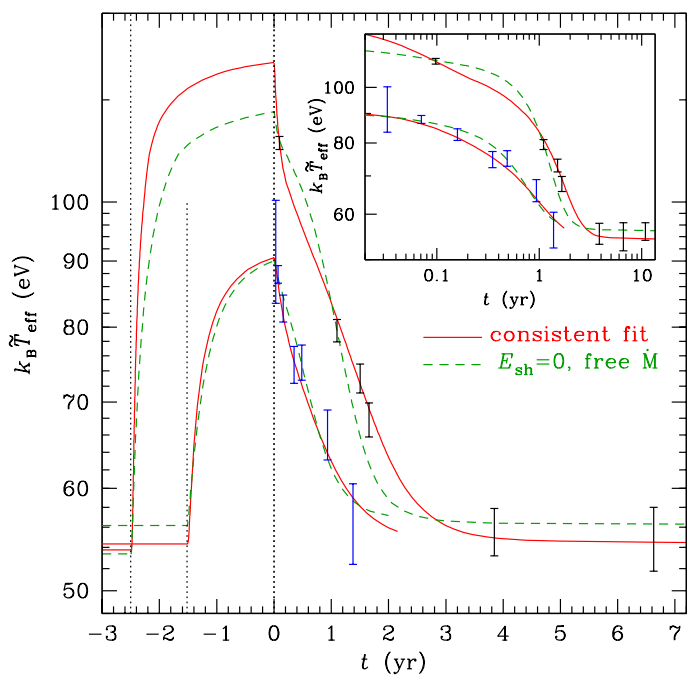}
\caption{Theoretical lightcurves and observations for the outbursts I and II
(upper and lower lines and errorbars, respectively). The dashed curve is
obtained without shallow heating, but with ad hoc adjusted accretion
rates (see text). The solid line reproduces the consistent fit (line 5
from Fig.~\ref{fig:MXB24}).
}
\label{fig:nonshallow}
\end{figure}
%%%%%%%%%%%%%%%%%%%%%%%%%%%%%%%%%%%%%%%%%%%%%%%%%%%%%%%%%

The last two features are illustrated in Fig.~\ref{fig:variations} for
the EoS BSk24 and outburst~I. We start from the
best-fitting curve in Fig.~\ref{fig:MXB24} (model 5 in
Table~\ref{tab:models}) and vary either the impurity parameter in the
accreted crust $Q_\mathrm{imp}$ (left panel) or the shallow heat per
baryon $E_\mathrm{sh}$ (right panel). The height and the average slope
of the thermal relaxation curve are nearly constant in the first and
second case and vary in the second and first case, respectively.
{Analogous trends were demonstrated in the
pioneering work by \citet{BrownCumming09}.}

Figure \ref{fig:variations} also allows us to roughly estimate the
magnitude of uncertainty of the adjustable parameters: the left panel
shows that $Q_\mathrm{imp}\approx3\pm1$, and the right panel shows that
$E_\mathrm{sh}\approx(230\pm30)$ keV, under the condition that the
other parameters of the theoretical model are kept fixed. The much
larger variations between different models in Table~\ref{tab:models}, as
well as in previous works (e.g., \citealt{Parikh_19}), are due to 
correlations of $Q_\mathrm{imp}$ and $E_\mathrm{sh}$ with other model
parameters.

Neglecting the above-mentioned first step of setting model
restrictions,  one can obtain an acceptable fit to the observed crust
cooling of MXB 1659$-$29 without shallow heating by varying other model
parameters, but at cost of disagreement with other observational data.
An example is shown in Fig.~\ref{fig:nonshallow}, where we compare the
best-fitting model~5 from Fig.~\ref{fig:MXB24} with another fit, where
we have set $E_\mathrm{sh}=0$ and let all accretion rates vary without
restrictions. In the latter case, we have obtained an acceptable
agreement with the observations using the model of a fully accreted
crust with helium envelope, $Q_\mathrm{imp}=0.5$,
$\langle\dot{M}\rangle=3\times10^{-11}\,M_\odot$
$\dot{M}=1.1\times10^{-8}\,M_\odot$ during the outbursts 0 and I, and
$\dot{M}=4.5\times10^{-9}\,M_\odot$ during the outburst II. Although the
crust cooling observations are satisfactorily described by this fit with
$E_\mathrm{sh}=0$, it is not acceptable, because it assumes a too low
mean accretion rate before outburst 0, too high accretion rates during
the outbursts, and an incorrect ratio between the accretion rates during
outbursts I and II.

%%%%%%%%%%%%%%%%%%%%%%%%%%%%%%%%%%
\subsection{Discussion}
\label{sect:discuss}

The self-consistent numerical models of the thermal evolution of
MXB1659$-$29 reproduce the equilibrium thermal luminosity as a result of
the long-term evolution, as well as the observed post-outburst decline
of the luminosity due to thermal relaxation of the crust for both
outbursts where this decline was accurately measured.

One of the results of our simulations is that the late-time decrease of
$\tTeff$ is absent not only in the traditional models of the inner crust
with low impurity content, but also in the models where deep layers of
the crust are composed of mixtures of different nuclei and have orders
of magnitude lower thermal conductivities than the traditional almost
pure inner crust. We traced the origin of this behavior to the small
heat capacity per baryon in the deep layers of the crust, which results
partly from the strong degeneracy and partly from the effect of neutron
superfluidity (Fig.~\ref{fig:cvtc}).

Previously, \citet{Deibel_17} succeeded in bringing theoretical crust
cooling to an approximate agreement with  
the possible late-time drop of thermal luminosity
(\citealt{Cackett_13}; the lowest errorbar in Figs.~\ref{fig:MXB24} and
\ref{fig:MXB25}) by assuming a high
impurity content of a bottom crust layer at $\rho>8\times10^{13}$ \gcc,
but only using a model where the heat capacity in this layer was
increased by 1\,--\,2 orders of magnitude due to the assumed closure of
the neutron superfluid gap at these densities. However, modern
theoretical models of the singlet-type neutron pairing 
\citep{MargueronSH08,Ding_16} converge to similar results and predict
that the neutrons remain superfluid at the considered densities and
temperatures. 

Some increase of the ion heat capacity may occur in the layers with high
$Q_\text{imp}$ due to a destruction of the long-range crystalline order
in a mixture of different nuclei. Then the impure bottom of the crust
may be in amorphous state. An increase of heat capacity in an amorphous
solid compared to a perfect crystal in the laboratory can reach a factor
of 1.5\,--\,2 (e.g., \citealt{Pohl81}, and references therein), but an
increase by orders of magnitude is not plausible. Moreover, the nuclei
are not the dominant contributors to the heat capacity near the crust
bottom (see Fig.~\ref{fig:cvtc}). Therefore the variations of the
effective temperature can hardly be explained by an increased heat
capacity of nuclei. 

Rapid late-time variations of soft X-ray flux during the post-outburst
neutron star crust cooling are not unique to MXB 1659$-$29. Recently,
\citet{Parikh_20} reported an unusually steep decay of $\sim7$ eV
followed by a rise of $\sim3$ eV in the observed effective temperature
during the crust cooling of two other SXTs, XTE J1701$-$462 and EXO
0748$-$676, around $\sim5.5$ years after the end of their outbursts.
Unlike the case of MXB 1659$-$29, these temperature variations are
difficult to explain by an increased hydrogen column density on the line
of sight. Among different tentative explanations discussed by
\citet{Parikh_20}, the most viable one appears to be convection, driven
by chemical separation at the ocean-crust boundary, which was previously
studied by \citet{MedinCumming14,MedinCumming15}. The latter authors
predicted dips of the effective temperature at $\sim5$\,--\,6 years of
crust cooling, similar to those observed for XTE J1701$-$462 and EXO
0748$-$676. We cannot exclude that the possible rapid fading of MXB
1659$-$29 about ten years after the outburst might have a similar
origin.

%%%%%%%%%%%%%%%%%%%%%%%%%%%%%%%%%%
\section{Conclusions}
\label{sect:concl}

We have studied the effects of crust composition and heating models on
thermal evolution of neutron stars in SXTs during and between outbursts.
We have shown that details of distribution of heating sources in the
crust can have an appreciable impact on the post-outburst cooling.
Our numerical results confirm the well-known expectations
that the nonstationary
thermal relaxation of the outer envelope is important for the light
curve in the first few months of the crust cooling.

We have demonstrated
that the deep layers of the inner crust can be composed of mixtures of
different nuclei and for this reason
can have relatively low electrical and thermal conductivities, without
invoking non-spherical nuclear shapes. We performed numerical
simulations of thermal evolution of a neutron star crust with such
mixtures during and after the outbursts and showed that the differences
between the cooling lightcurves with and without such impure layers can
be appreciable, but unlikely very large. 

To test our theoretical models against observations, we performed
self-consistent simulations of the long-term and short-term thermal
evolution of the transiently accreting neutron star MXB 1659$-$29 using
different models of crust composition. As in the previous works, we
find that the ``shallow heating'' in addition to the deep crustal
heating is necessary to reproduce the early-time post-outburst thermal
luminosities. We found that the presence of the highly impure deep
layers in the inner crust is not crucial for comparison of the
theoretical crust cooling with observations. However, the needed shallow
heat depends on the assumed composition of the crust and heat-blanketing
envelope. More transparent
outer envelope models require weaker shallow
heating, which is easier to  agree with the theoretical constraints
recently obtained by \citet{Chamel_20}.
On the other hand, the enhanced
transparency of the envelope requires an increased impurity content in
the crust to reproduce the observed crust relaxation at the timescale of
years.

In this work we have used the models by \citet{HZ08} and
\citet{Fantina_18} for the deep crustal heating. Recently,
\citet{ChugunovShchechilin20} and \citet{GusakovChugunov20} have shown
that diffusion of neutrons in the inner crust, neglected in the previous
studies, can be essential for composition, equation of state, and heating
of the accreted inner crust. An impact of these effects on thermal
evolution of transiently accreting neutron stars remains to be studied
in a future work.

\begin{acknowledgements}

The work of A.P.{} was supported  by the
Russian Science Foundation (grant 19-12-00133).

\end{acknowledgements}

\newcommand{\arnps}{Annu.\ Rev.\ Nucl.\ Part.\ Sci.}
\newcommand{\al}{Astron.\ Lett.}
\newcommand{\jaa}{JA\&A}
\newcommand{\rmp}{Rev.\ Mod.\ Phys.}

\end{document}